\documentclass[twocolumn,aps,amsmath,amssymb,superscriptaddress]{revtex4}

\usepackage{graphics}
\usepackage{graphicx}
\usepackage{dcolumn}
\usepackage{bm}
\usepackage{color}
\usepackage{soul}
\usepackage{textcomp}
\usepackage{multirow}
\usepackage{setspace}
\usepackage[mathlines]{lineno}
\usepackage{textcomp}
\usepackage{mathcomp}
\usepackage{spverbatim}
\usepackage{float}
\usepackage{longtable}
\usepackage{amssymb}
\usepackage{pifont}
\usepackage{amsmath}
\usepackage[bb=dsserif]{mathalpha}
\usepackage{bm}
\usepackage{soul}

\usepackage[T1]{fontenc}       
\usepackage[utf8]{inputenc}
\usepackage{amsmath,amsfonts}
\usepackage{epsfig}
\usepackage{rotating}
\usepackage{longtable}
\usepackage{dcolumn}
\usepackage{epstopdf}
\usepackage{pdfpages}
\usepackage{MnSymbol}
\usepackage{xcolor}

\usepackage{soul}

\usepackage{hyperref}
\hypersetup{
    colorlinks=true,       
    linkcolor=blue,        
    citecolor=blue,        
    filecolor=blue,        
    urlcolor=blue,         
    runcolor=blue
}

\newcommand{\ket}[1]{\left| #1 \right\rangle}
\newcommand{\bra}[1]{\left\langle #1 \right|}
\newcommand{\braket}[2]{\left\langle #1 \right|\left. #2 \right\rangle}

\newcommand{\UDEA}{Grupo de F\'isica At\'omica y Molecular, Instituto de F\'isica, Universidad de Antioquia, Medell\'in, Colombia.}
\newcommand{\UCN}{Department of Physics, Universidad Cat\'olica del Norte, Av. Angamos 0610, Antofagasta, Chile.}

\begin{document}


\title{Ultrafast non-adiabatic molecular energy conversion into photons induced by quantized electromagnetic fields}


\author{Arley Flórez López}
\affiliation{\UDEA}
\author{Johan F. Triana}
\email{johan.triana@ucn.cl}
\affiliation{\UCN}
\author{Jos\'e Luis Sanz-Vicario}
\email{jose.sanz@udea.edu.co}
\affiliation{\UDEA}

%
%

\date{\today}

\begin{abstract}
Molecular polaritons within the mid-infrared regime have emerged as a source for modifying and manipulating molecular and photonic properties. However, the development of new methodologies for photon generation is still a challenge in nanophotonics. We propose a molecular model based on the Holstein-quantum-Rabi Hamiltonian, which also incorporates realistic dipole moments and non-adiabatic couplings among electronic excited states, to study the ultrafast photodynamics of diatomic molecules in confined electromagnetic fields within quantized cavities. In addition to vibronic transitions due to intrinsic non-adiabatic couplings, two types of light-induced crossings emerge: one type is located at molecular nuclear geometries where the rotating wave approximation is fulfilled, and another type appears at different geometries where counter-rotating transitions may occur. 
We make a comprehensive study of polariton photodynamics within a time window of a few tens of femtoseconds, where dissipative mechanisms do not influence the polariton photodynamics. 
We stress the dramatic change of the polariton energy spectrum as a function of the Huang-Rhys factor when non-adiabatic couplings are included in the model. We conclude that both the molecular non-adiabatic couplings and, more specifically, the counter-rotating couplings in the cavity-molecule interaction play a crucial role in converting vibronic energy into photons through excited dressed states. We also show that the sign of the Huang-Rhys factor has a significant impact on this photon conversion. Our work paves the way for the development of many-photon generation powered by strong light-matter interaction, along with potential applications using alkaline earth monohydride molecules. 
\end{abstract}

\maketitle


\section{Introduction}
\label{sec:intro}

A full \textit{ab initio} computation of the photodynamics of molecular polaritons may become a formidable problem because of the number of degrees of freedom that quantized radiation incorporates into the matter system \cite{fregoni2022, Xiang2024, Weight2025}. 
Molecular aggregates, which are normally present in extant experiments, add even more complexity to the solution \cite{Spano2010,Herrera2014,Herrera2016,  Hestand2018}. 
There has already been interest in full \textit{ab initio} proposals that combine quantum chemistry methods and quantum optics tools \cite{Triana2018,Szidarovszky2021,Sidler2023,Ruggenthaler2023}, although limited to single molecules or a small ensemble of molecules, very far from any thermodynamic limit.  
It is astonishing how simple models in quantum optics, like the Jaynes-Cummings model \cite{Jaynes1963,Gerry2004,Larson2024} for a single two-state system coupled to quantized radiation and the corresponding Tavis-Cummings model \cite{Tavis1968,garraway2011}  for a collective of identical two-state systems [both in the rotating wave approximation (RWA)] have served well to predict and understand many physical phenomena involved in the strong-coupling radiation-matter interaction, e.g., single photon generation using quantum emitters \cite{Chen2022}. 
These models and extensions have been widely applied to simple molecules in strong vibrational coupling due to cavity modes in the infrared frequency \cite{Herrera2016,Herrera2018,Ribeiro2018,Hernandez2019,Triana2020,Schnappinger2024}. 
More recently, they have been applied to molecules in the strong-coupling regime with photons in the visible and UV region, leading to electronic excitation \cite{Kowalewski2016,Triana2018,Triana2019,Weight2025}.
A photo-excitation restricted to the Franck-Condon region with the involvement of at most two molecular states (or, for simplicity, the HOMO-LUMO frontier orbitals) makes the application of these models in molecules potentially successful. For example, the collective Tavis-Cummings model has recently been used to perform a detailed study of aggregates embedded in optical cavities through coherent two-dimensional (photon-echo) spectroscopy (2DS), with a remarkable explanation of the experimental findings in terms of temporal emergence and evolution of diagonal and cross peaks in the 2D spectra \cite{Gallego2024}. However, in the latter work, while the electronic degree of freedom was accounted for, the vibrations were not included explicitly, and instead the structure and dynamics of molecular phonons were introduced as part of the enclosing bath in an open quantum system approach, which eventually plays the role of a fast dissipative vibrational decay (within the electronic states) due to dephasing terms. 

To explicitly include, from first principles, exciton, vibrational, and photon degrees of freedom in simulations of many-molecule polaritonics, plus a realistic bath consisting of an ensemble of harmonic or anharmonic oscillators, is out of today's computational capabilities. However, we believe that there is still room to implement simple molecular models to gain insight into the fast polariton photodynamics. 
To introduce molecular models close to the realm of photodynamics within excited states, we must account for \textit{i)} the non-adiabatic couplings (NAC) around avoided crossings in 1D (conical intersections in higher dimensions) and \textit{ii)} the emergent light-induced crossings (LIC)  caused by the photonic cavity modes \cite{Szidarovsky2018, Triana2018}. 
Both interactions, via matter non-adiabatic and light-matter dipole couplings, must depend on the molecular geometry.

We show that every molecule exhibits two types of light-induced crossings around avoided crossings in diatoms or conical intersections for polyatomic molecules. 
The first type corresponds to rotating wave approximation (RWA) transitions and the second type to non-RWA transitions. Both induced non-adiabatic transitions become accessible for an initial vibronic excitation that carries a large energy excess. 
This extra energy allows for photon generation by redistributing energy among the electronic, vibrational, and photonic degrees of freedom when molecules initially in excited electronic states interact with the confined electromagnetic vacuum in nanocavities.
In this regard, previous studies have shown photon generation in 2D materials \cite{Vogl2019,Gupta2023}, microwave cavities \cite{Wilson2010}, quantum dots coupled to plasmonic nanocavities \cite{Hoang2016,Dusanowski2020, Wu2022} and through isomerization processes \cite{Perez2020}. 

To study the collaborative effect of natural and induced non-adiabatic transitions, we introduce an extended Holstein-quantum-Rabi (HQR) model for diatoms coupled to quantized radiation by adding non-adiabatic effects in a realistic manner. We add the dependence on the molecular geometry in \textit{i)} molecular non-adiabatic couplings, when potential energy curves display avoided crossings and \textit{ii)} transition dipole moments. 
Our proposed HQR model allows us to implement localized interactions to better understand the role of each non-adiabatic coupling during the early stages of molecular polariton photodynamics. 
We have not included any dissipative effects due to the cavity, to the molecules, or to the environment, so that we follow the ultrafast unitary evolution in the first tens of femtoseconds. Our results and conclusions remain valid before the onset of cavity photon loss, molecular dephasing, etc.      

The paper is organized as follows. The concept and equations for the non-adiabatic Holstein-quantum-Rabi model are introduced in Sec. \ref{sec:methods}, that comprises the detailed description of the Hamiltonian, which follows from a molecular diabatization procedure and the modeling of the non-adiabatic cavity and electrostatic interactions. Details on the construction of the Hamiltonian are relegated to  appendices \ref{sec:AppendixA} and \ref{sec:AppendixB}.
We also introduce the well-known LICs, specifically classified into rotating LICs and counter-rotating LICs, where the RWA only applies to the former case. 
For the numerical solution of photodynamics, we use two different methods that yield the same results upon convergence: \textit{i)} spectral expansions in terms of tensor products of photon, exciton and phonon states, $\ket{n_{\mathrm{c}};g/e,n}$ (using QuTiP libraries), and \textit{ii)} the time-dependent multiconfigurational time-dependent Hartree (MCTDH) method, with a spatial representation of the quantized radiation. For concreteness, we propose studying the polariton photodynamics of an alkaline-earth monohydride molecule, CaH, due to the particular arrangement of the electronic ground and the two lowest excited states, all with the same symmetry $^2\Sigma^{+}$, and characterized by a positive Huang-Rhys factor between the two lowest excited states.     
Section \ref{sec:results} contains the results and discussion on polariton dynamics. We study the performance of our HQR model fitted to CaH molecular parameters and discuss the underlying mechanisms based on the well-localized interactions of LICs and NAC. 
A temporal trapping of photons dressing the matter is described, as well as the asymmetry in photon conversion against the variation of the Huang-Rhys factor. Finally, results on polariton photodynamics in a realistic {\em ab initio} calculation in CaH are presented and discussed with the tools already learned from the extended HQR model. We end up in Sec. \ref{sec:conclusions} with conclusions and perspectives derived from this work.

\section{\label{sec:methods} Theory}

 We introduce a molecular model for a single diatomic molecule that is strongly coupled with quantized radiation, including degrees of freedom for electrons, vibrations, and photons.
 To better approximate the photodynamics exhibited by real molecules in electronic excited states, we explicitly introduce the transition dipole moment and the non-adiabatic couplings that depend on the internuclear distance $q$.
 The importance of a $q$-dependent dipole is due to the creation of the so-called cavity light-induced crossings (LIC) at distances where photon-dressed potential energy curves intersect \cite{Kowalewski2016,Triana2018,Szidarovsky2018,Szidarovszky2021,Fabri2022}. 
Both the $q$-position and energy-location of these cavity LICs depend not only on the shape of the coupled molecular potential energy curves but also on the cavity mode frequency $\omega_{\rm c}$. The choice of a kind of molecules or a set of different cavity mode frequencies offers some flexibility in manipulating the molecular photodynamics \cite{Herrera2020}. 

Light-induced non-adiabatic transitions are eventually produced if the magnitude of the dipole moment is relevant at the LIC position.
All molecules present a very complex landscape in the manifold of excited electronic states, and their photochemistry strongly depends on the paths followed by the excited wave packet before decaying to the ground state. 
The presence of avoided crossings among excited states in diatoms or conical intersections in many-atom molecules is the norm more than the exception, and they must be accounted for in any realistic molecular simulation. 
We have included these two competing non-adiabatic effects in our molecular model. 

Our Holstein-quantum-Rabi model makes use of harmonic potentials for the two excited states, which have the same symmetry, i.e., they do cross to each other instead of displaying an avoided crossing according to the Wigner-von Neumann rule \cite{Wigner1929,Landau1991}.
This implies a preliminary step; better than the adiabatic representation (avoided crossings with non-adiabatic couplings), we adopt a diabatic representation (real crossings with residual electrostatic potential couplings).
The latter couplings are due to the electronic Hamiltonian itself, as the diabatic states are no longer eigenstates of the Hamiltonian \cite{Smith1969}. 
The transition from a non-adiabatic picture to a diabatic picture (performed through a unitary transformation) is widely used, mainly because diabatic couplings are softer and better manageable than sharp non-adiabatic couplings \cite{Bransden1992}.
 
We are interested in the first steps of the fast phototodynamics (in a few fs time window), and dissipative effects (photon loss, molecular dephasing, etc.) are not considered. 
We highlight that our model involves molecular electronic excitations within the infrared regime, the region of the electromagnetic spectrum where structured nanocavities can operate with photon lifetimes on the order of hundreds of femtoseconds \cite{Muller2018, Ahn2023, Simpkins2023}.
Atomic units are used unless otherwise stated.

\subsection{\label{sec:holstein} Holstein-quantum-Rabi model for a diatomic molecule}

The fast photodynamics of diatoms within quantized cavities can be modeled using a Hamiltonian approach that incorporates electronic (excitons), vibrational (phonons), and quantized radiation (photons) degrees of freedom. Molecular rotation is always much slower than the previous modes and is therefore not considered.  
Any molecular electronic potential energy curve admits a Taylor expansion to the second order around the equilibrium distance $q_x$ in the form $V(q)= V(q_x) + d^2 V(q)/dq^2|_{q_x} (q-q_x)^2/2$; it justifies the use of harmonic states for the lowest bound states. Our proposed molecular Hamiltonian contains two electronic states $x=g,e$ and corresponds to a diabatic- and dipole-coupled Holstein-quantum-Rabi model in the form 
(see details in the Appendices \ref{sec:AppendixA} and \ref{sec:AppendixB})
\begin{equation}
\hat{H}=\hat{H}_{\textrm{mol}} + \hat{H}_{\textrm{rad}} + \hat{H}_{\textrm{int}},
\label{eq:hamiltonian}
\end{equation}
where
    \begin{equation}
        \begin{split}
            & \hat{H}_{\textrm{mol}} = \omega_g \left( \hat{b}^{\dagger}\hat{b}+\frac{1}{2} \right) | g\rangle \langle g| \\ 
            & + \left\{ \omega_{ge} + \omega_e \left[\hat{D}(\lambda)\,\hat{S}(r)\,\hat{b}^{\dagger}\hat{b}\,\hat{S}^{\dagger}(r)\hat{D}^{\dagger}(\lambda) + \frac{1}{2} \right] \right\}| e\rangle \langle e| \\ 
            & + V\,\exp \left\{ -\frac{(\hat{q}-q_{\mathrm{c}})^2}{2\sigma^2}\right\} (\hat{\sigma}^{+}+\hat{\sigma}^{-}),
        \end{split}
        \label{eq:hamiltonianMol}
    \end{equation}
represents the diabatically coupled two-state molecular Hamiltonian for states $|g\rangle$ and $|e\rangle$, with different harmonic frequencies $\omega_g$ and $\omega_e$ and different equilibrium distances $q_g$ and $q_e$ (related through the Huang-Rhys factor $\lambda$), respectively,  with a vertical energy separation between the two potential energy curves given by $\omega_{ge}$ 
(see Appendix \ref{sec:AppendixA}).
Here, $\hat{\sigma}^+ = | e \rangle \langle g |$, $\hat{\sigma}^- = | g \rangle \langle e |$ are the (fermionic) exciton operators and $\hat{b}^\dag$ and $\hat{b}$ are the (bosonic) vibration operators. 
The shift and shape of the potential energy curve of the state $\ket{e}$ is provided by the displacement operator $\hat{D}(\lambda)$ with the Huang-Rhys factor $\lambda=q_e\sqrt{M \omega_g/2}$ and the squeezing operator $\hat{S}(r)$ with a factor $r=\log(\sqrt{\omega_e/\omega_g})$, where $M$ is the reduced mass of the system. 

\begin{figure}[t]
    \centering \includegraphics[width=0.95\linewidth]{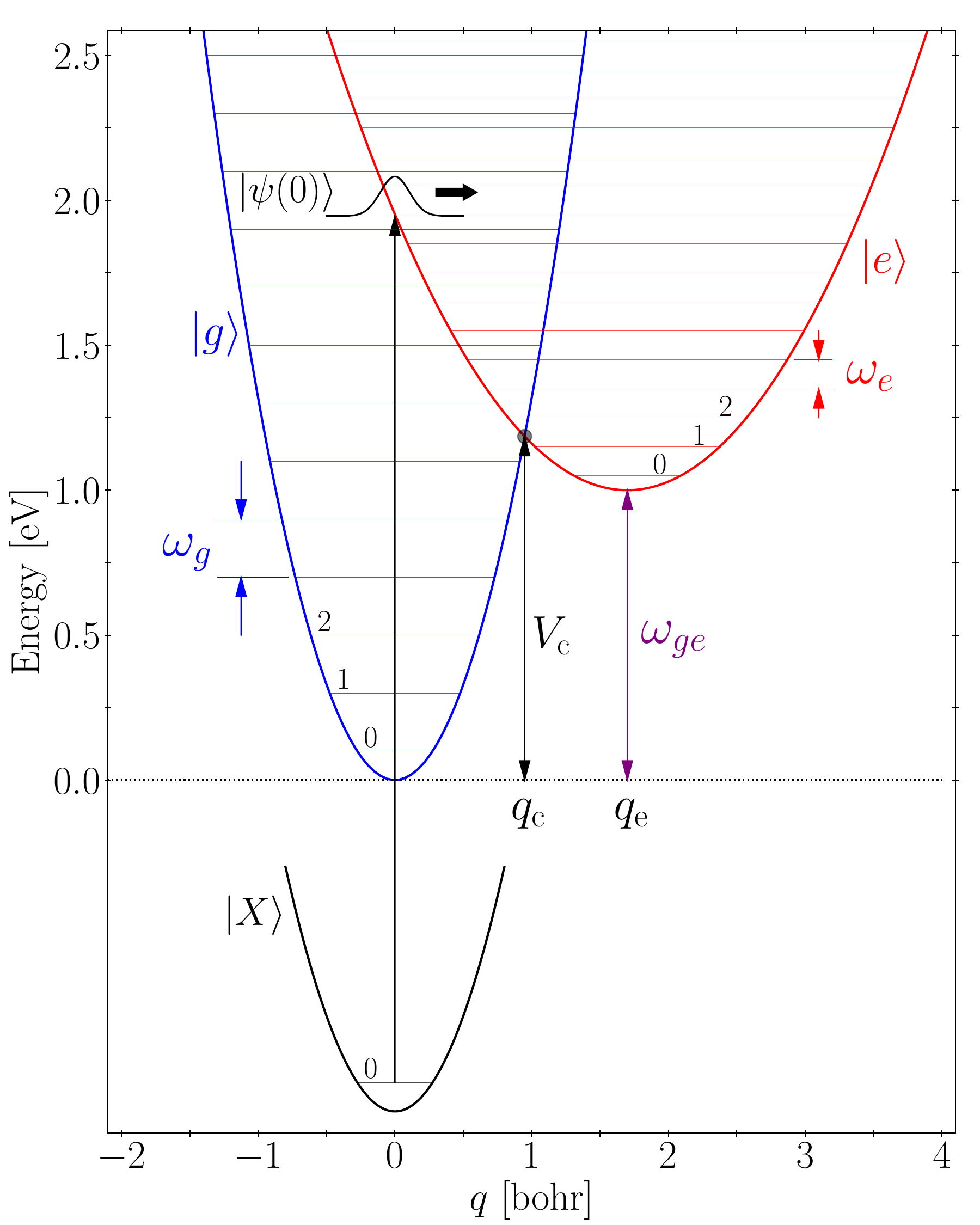}
    \caption{ Energy scheme for the Holstein-quantum-Rabi molecular model. The electronic model is composed by two
    excited states $|g\rangle$ an $|e\rangle$ whose potential energy curves $V_g(q)$ and $V_e(q)$ are modeled as 
    harmonic oscillators with frequencies  $\omega_{\mathrm{g}}=0.2$ eV and $\omega_{\mathrm{e}}=0.1$ eV and equilibrium distances
    $q_g=0$ and $q_e=1.7$ bohr (Huang-Rhys factor $\lambda=4.37$), respectively. The chosen energy difference between the potential 
    energy curves is $\omega_{\mathrm{ge}}=1.0$ eV. The crossing point between the two curves is located at $q_{\mathrm{c}}=0.95$ bohr. 
    The photoabsorption process starts from the vibrational ground state of the electronic ground state $|X\rangle$ and the photodynamics
    in the manifold of excited states starts from a vertical Frank-Condon excitation $|X\rangle \to |e\rangle$ with initial state $|\Psi(0) \rangle$ and initial energy $E_0 \sim 2$ eV.}
    \label{fig:figure1}
\end{figure}

The last term in Eq.~\eqref{eq:hamiltonianMol} is responsible for non-adiabatic transitions between electronic states $|g\rangle$ and $|e\rangle$. 
This coupling represents the residual electrostatic coupling emerging after the diabatization procedure from the two adiabatic PECs that show an avoided crossing at the internuclear distance $q_{\mathrm{c}}$. 
Diabatization is currently performed in most molecular calculations to avoid the introduction of sharp non-adiabatic couplings at avoided crossings. Since this model relies on harmonic potentials that cross each other, we choose to work on the diabatic picture.
We simulate the diabatic couplings with a Gaussian of width $\sigma$ centered at the crossing point $q_{\mathrm{c}}$ between the two potential energy curves. 
This crossing point depends on the choice of the Huang-Rhys factor $\lambda$ and their expressions are 
\begin{equation}
q_{\mathrm{c}}(\lambda) = q_e(\lambda)/2 + \omega_{ge}/M q_e(\lambda) \omega_g^2; r=0, 
\end{equation}
and 
\begin{equation}
q_{\mathrm{c}}(r,\,\lambda) = \frac{q_e(\lambda)}{1-e^{-4r}} \left[ 1 - \sqrt{ e^{-4r} +  \frac{2 \omega_{ge} ( e^{-4r} - 1 )} {M \omega_e^2 q_e^2(\lambda) }} \right]; r \ne 0.     
\end{equation}
The corresponding energy at the crossing point is $V_{\mathrm{c}}(\lambda)=M \omega_g^2 q_{\mathrm{c}}(\lambda)^2/2$. 
This molecular model is depicted graphically in Fig.~\ref{fig:figure1} for the parameters specified in the figure caption. 

As plotted in Fig.~\ref{fig:figure1}, we assume that a prior vertical (Frank-Condon) laser-induced photoexcitation occurs from the ground state $\ket{X}$ (not explicitly included in the model Hamiltonian) to the excited state $\ket{e}$, which brings an initial molecular wave packet $\ket{\Psi(t=0)}$ to the inner turning point of the PEC $V_e(q)$, and subsequently the wave packet moves rightward subject to a molecular diabatic coupling located at the crossing point $q_{\mathrm{c}}$. 
Due to the (diabatized) non-adiabatic coupling, here called DC, the wave packet splits into two components, one that keeps on moving in the PEC $V_e(q)$ and another that is transferred to move in the PEC $V_g(q)$. 
From that crossing point $q_{\mathrm{c}}$, two wave packets emerge moving with approximate periods $T_g= 2\pi/\omega_g$ and $T_e=2\pi/\omega_e$. 
For the case in Fig.~\ref{fig:figure1}, the chosen frequencies are $\omega_g=0.2$ and $\omega_e=0.1$ eV; when the wave packet portion in the $\ket{e}$ state ends a full cycle, the wave packet component in the $\ket{g}$ state makes two cycles, indicatng that the latter wave packet visits the NAC coupling twice the times that the wave packet in the $\ket{e}$ state does. 
These different temporal periods for the two wave packets will become relevant in the forthcoming analysis. Now, in Eq. ~\eqref{eq:hamiltonian},
\begin{equation}
        \hat{H}_{\rm rad}=\omega_{\rm c} \hat{a}^{\dagger}\hat{a}
    \end{equation}
is the Hamiltonian for the quantized radiation of the photonic mode, with (boson) photon operators $\hat{a}^\dag$ and $\hat{a}$, and
\begin{equation}
        \hat{H}_{\rm int}= g\,\exp \left\{-\frac{(\hat{q}-q_{\mathrm{c}})^2}{2\beta^2} \right\}(\hat{\sigma}^{+}+\hat{\sigma}^{-}) (\hat{a}^{\dagger} + \hat{a})
        \label{eq:Cavityint}
    \end{equation}
is the molecule-cavity interaction with a dipole coupling factor 
$g=\chi d_0\sqrt{\omega_{\mathrm{c}}/2}$ that contains the cavity strength $\chi$, the maximum of the molecular transition dipole moment $d_0$ and the cavity mode frequency $\omega_{\rm c}$. 

The interaction term in Eq.~\eqref{eq:Cavityint} contains both the rotating R (RWA and energy-preserving) and counter-rotating CR (energy non-conserving) terms. The dipole coupling with factor $g$ is modeled with a Gaussian distribution with a width $\beta$ centered at the crossing point $q_{\mathrm{c}}$ between the two curves.  
This model agrees with realistic dipole moments since most heteronuclear diatomic molecules exhibit extended dipole moments along the internuclear distance, mostly enclosing the crossing point, which is the relevant region for the photodynamics.   
Hence, the Hamiltonian in Eq.~\eqref{eq:hamiltonian} contains most of the ingredients present in molecule-cavity non-adiabatic photodynamics involving electronic excited states of diatomic molecules, and explicitly includes their vibrational structure.
\begin{figure}[t]
    \centering
    \includegraphics[width=0.95\linewidth]{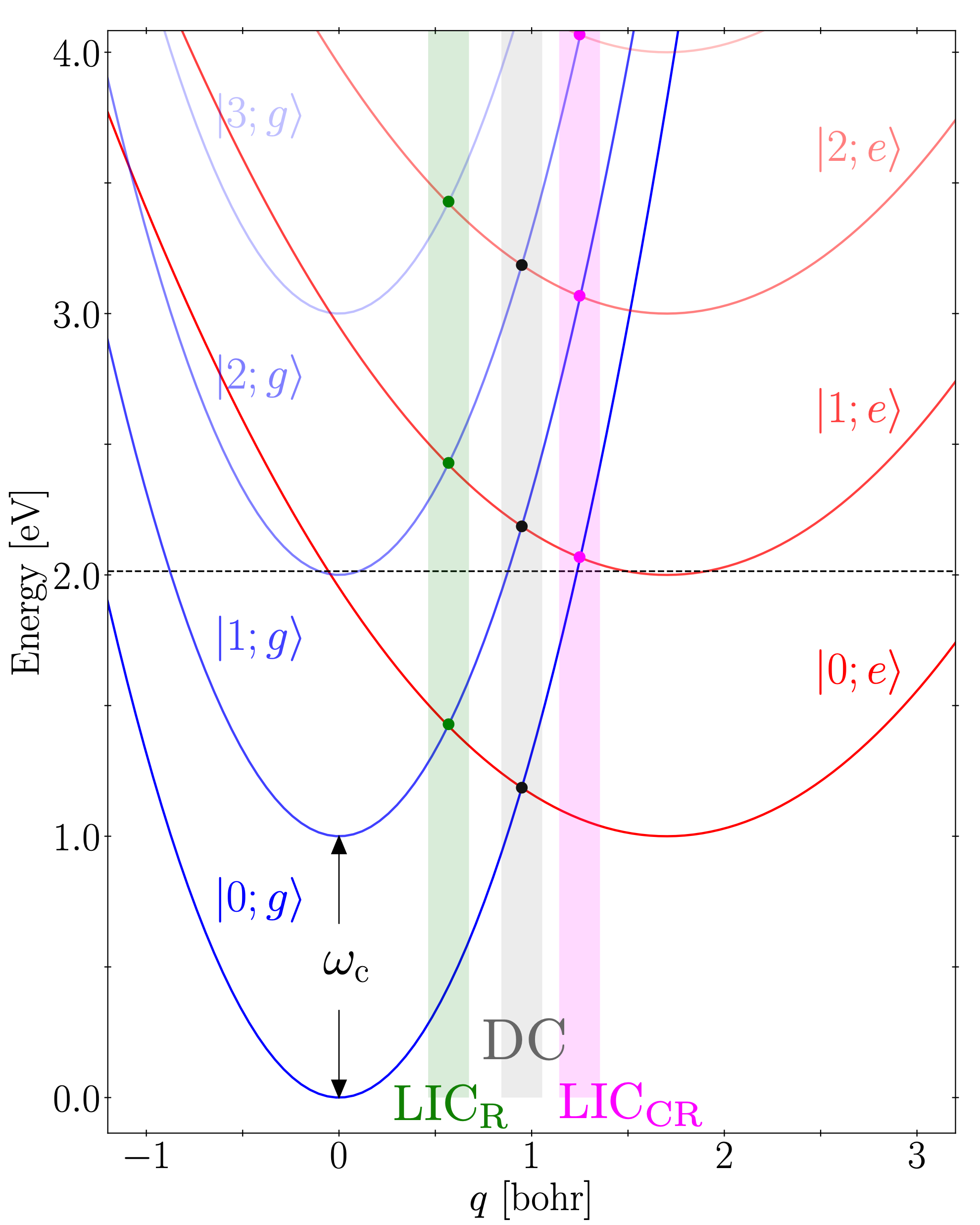}
    \caption{Cavity field dressed states $\{ | n_{\mathrm{c}}; g\rangle, | n_{\mathrm{c}}; e \rangle \}$ of the excited states $|g\rangle$ and $|e\rangle$ in Fig. \ref{fig:figure1} 
    for a number of cavity photons $n_{\mathrm{c}}=0,1,2$ and 3, with a resonant photon energy  $\omega_{\mathrm{c}}=\omega_{\mathrm{ge}}$.
    The different types of crossings are denoted by solid circles: a set of diabatic crossings (DC) located at $q_{\mathrm{c}}=0.95$ bohr (black circles)
    rotating (RWA) light-induced crossings (LIC$_{\mathrm{R}}$) at $q < q_{\mathrm{c}}$ and counter-rotating (non-RWA) light-induced crossings (LIC$_{\mathrm{CR}}$) at
    $q > q_{\mathrm{c}}$. 
    }
    \label{fig:figure2}
\end{figure}

To understand the molecular polariton structure and photodynamics, it is important to visualize and interpret the molecular diagrams corresponding to the light-dressed PECs.
These PECs are obtained from those electronic states $|g\rangle$ and $|e\rangle$ in Fig. \ref{fig:figure1} by adding the energy of $n$ cavity photons with a frequency mode $\omega_{\mathrm{c}}$ to produce states $|n_{\mathrm{c}}; g\rangle$ and $|n_{\mathrm{c}}; e \rangle$, with $n_{\mathrm{c}}=0,1,2,3,...$, as shown in Fig. \ref{fig:figure2}. 
In the figure we indicate the presence of a series of crossings, namely, \textit{i)} Light-induced crossings due to {\em rotating} terms ($\hat{a} \hat{\sigma}^+, \hat{a}^\dag \hat{\sigma}^-$) in the molecule-field interaction that we call LIC$_{\mathrm{R}}$ and located before the diabatic crossings (DC) inherent to the molecule itself; \textit{ii)} Light-induced crossings due to {\rm counter-rotating} terms  ($\hat{a}^\dag \hat{\sigma}^+, \hat{a} \hat{\sigma}^-$), that we call here LIC$_{\mathrm{CR}}$, and located after the DC.  

Light-induced crossings due to rotating terms (LIC$_{\mathrm{R}}$) only connect pairs of states $ \{ |n_{\mathrm{c}}+1; g \rangle, |n_{\mathrm{c}}, e\rangle \}$ for $n_{\mathrm{c}}=0,1,2,3,...$ in the same excitation manifold (the same eigenvalue for operator $\hat{N}_e = \hat{a}^\dag{a} + \hat{\sigma}^+ \hat{\sigma}^-$) so that LIC$_{\mathrm{R}}$ is related to the RWA, already implemented in the Jaynes-Cummings model. 
In addition to LIC$_{\mathrm{R}}$ and DC, molecules also present the other kind of crossings, LIC$_{\mathrm{CR}}$, that connect pairs of states $ \{ |n_{\mathrm{c}}; g \rangle, |n_{\mathrm{c}}+1, e\rangle \}$ for $n_{\mathrm{c}}=0,1,2,3...$ that belong to different excitation manifolds, thus involving also non-RWA transitions present in a quantum Rabi model. 
Therefore, unlike atoms, molecules intrinsically display both types of radiation crossings, LIC$_{\mathrm{R}}$ and LIC$_{\mathrm{CR}}$, separated along the internuclear distance, which means that they eventually act separately when we consider wave packet dynamics in the excited states. 
It should be noted that while the position of the DC remains always unchanged, the $q-$position for the LIC$_{\mathrm{R}}$ and the LIC$_{\mathrm{CR}}$ can be selected by a judicious choice of the cavity mode frequency $\omega_{\mathrm{c}}$. 
For small values of $\omega_{\mathrm{c}}$, both LICs approach the DC, and for higher values of $\omega_{\mathrm{c}}$, both LICs separate from the position of the DC. This implies that the distinct physical features that emerge from each crossing could eventually be isolated. 

Other factors are the separation between the excited states $\ket{g}$ and $\ket{e}$, vertically, given by the quantity $\omega_{ge}$ and horizontally, given by the Huang-Rhys factor $\lambda$, as explained above.
The variation of $\lambda$ can only be achieved by selecting different molecules. Molecules show different arrangements in the manifold of excited states; some extend their internuclear distance when electronically excited ($\lambda > 0$, as in our figures) and others contract it upon electronic excitation ($\lambda < 0$). We demonstrate in the following that polariton photodynamics in molecules with a positive Huang-Rhys factor produces different yields than in molecules with a negative Huang-Rhys factor. 
This could be naively unexpected since the situations for $\pm \lambda$ are connected through a mirror symmetry (a plane placed at $q=0$) for the $\ket{e}$ state (see Fig. \ref{fig:figure2}). The fact that breaks the symmetry is that the ground state PEC $V_{\mathrm{X}}(q)$ (which provides the initial state) is clamped to that of the excited state $V_e(q)$ so that the distance $|q_e-q_X|$ remains unchanged to maintain the same initial total energy $E_0$ after a Frank-Condon vertical excitation.
We realize from Fig.~\ref{fig:figure1} and Fig.~\ref{fig:figure2} that only a limited number of uncoupled states $|n_{\mathrm{c}}; x, n \rangle$ for $x=g,e$ may be involved in the spectral expansion of the total polariton state. 
In principle, only 10 vibrational states of $| 0;g \rangle$, 5 states of $\ket{1;g}$ and 10 states for $\ket{0;e}$ lie below 2 eV. In addition, one LIC$_{\mathrm{R}}$ and one DC crossing would be mainly involved in the photodynamics (see Fig.~\ref{fig:figure2}). 
 
\begin{figure}[t]
    \centering   \includegraphics[width=0.95\linewidth]{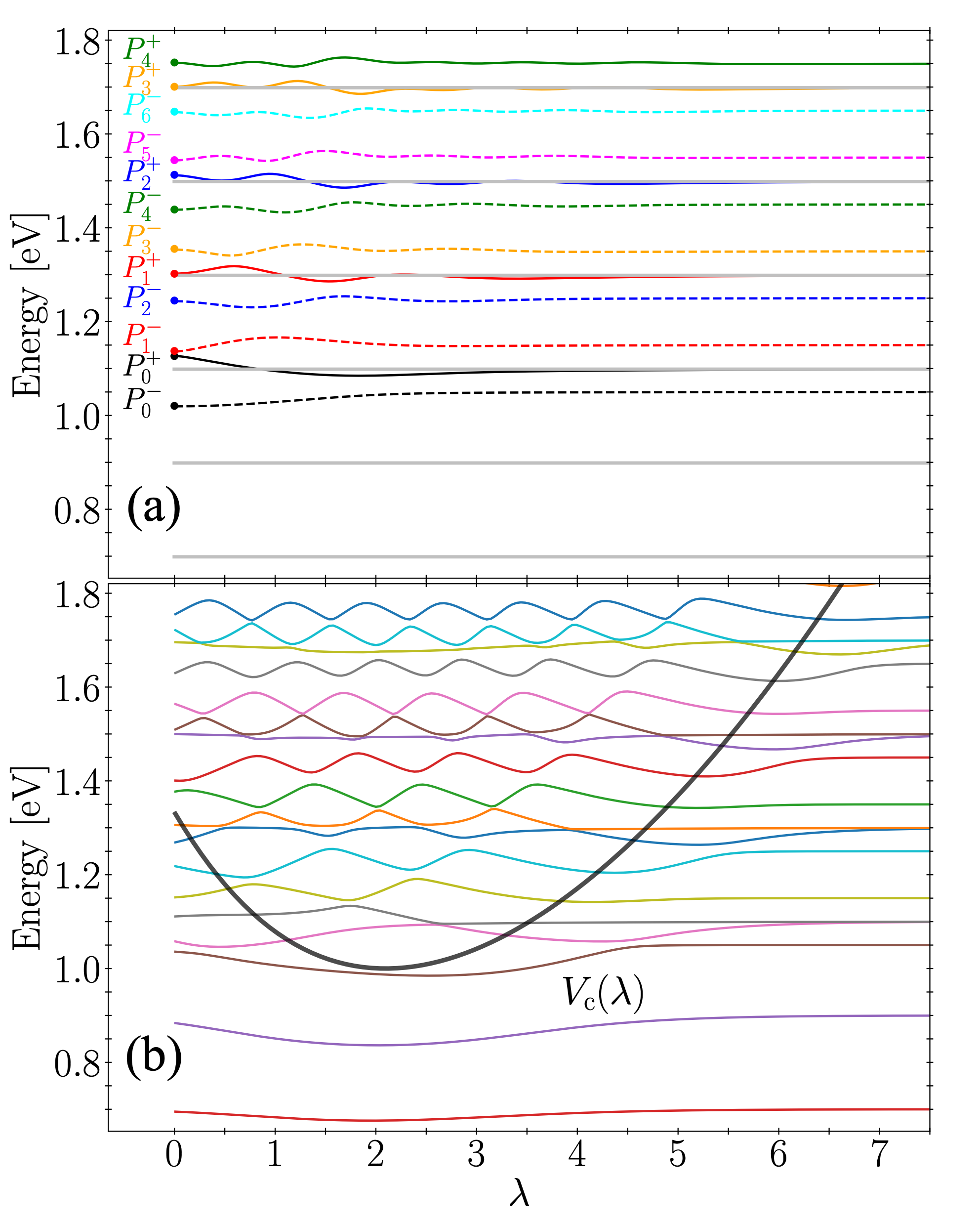}
    \caption{Vibronic polariton energies of the Holstein-quantum Rabi model within the set of  
    uncoupled states $|0; g, n\rangle$,  $|1; g,n' \rangle$ and $|0; e, n'' \rangle$ against the variation of the Huang-Rhys factor $\lambda$ 
    (a) with a constant electronic dipole coupling $d(q)=g$ and without diabatic couplings $V_{\mathrm{D}}(q)=0$.  
    Eigenstates are denoted as $P^{-}_n$ ($\lambda=0$) for the lower vibronic polariton states (dashed lines) and $P^{+}_n$ for the upper vibronic polariton states (solid lines)
    for vibrational numbers $n=0$ (black), 1 (red), 2 (blue), 3 (yellow) and 4 (green), as a result of the cavity coupling between $|1; g,n \rangle$ and $|0; e, n \rangle$ vibronic states. 
    The five vibronic states $|0; g,n \rangle$ included in the figure (solid grey lines) are uncoupled states of the ground state,
    (b) with a $q-$dependent dipole coupling  $d(q)$ and a $q-$dependent diabatic coupling $V_{\mathrm{D}}(q)$ (see text). $V_c (\lambda)$ indicates the energy value at the position of the crossing $V(q_c)$ for each $\lambda$. Molecular parameters as in Fig. \ref{fig:figure1}.}
    \label{fig:figure3}
\end{figure}

\subsection{Molecular polariton eigenstates of HQR model}
\label{subsection:Poleigenstates}

The vibronic polariton states, i.e., eigenstates of the Holstein-quantum-Rabi model, can be obtained by diagonalization of the Hamiltonian in Eq. ~\eqref{eq:hamiltonian} represented in terms of an uncoupled basis $|n_{\mathrm{c}}; g, n' \rangle$ and $|n_{\mathrm{c}}; e, n'' \rangle$. 
As an example, we build a system similar to the case previously studied in \cite{Calvo2020}, where it was assumed that the dipole interaction term $g (\hat{\sigma}^{+}+\hat{\sigma}^{-}) (\hat{a}^{\dagger} + \hat{a})$ contains a constant strength $g$ (no dependence on the internuclear distance $q$). 
We choose similar parameters already used in Ref.~\cite{Calvo2020} for comparison, but in our case for two different harmonic frequencies. 
Therefore, the harmonic frequencies are those used to plot Figs. \ref{fig:figure1} and \ref{fig:figure2}, $\omega_g=0.2$ eV and $\omega_e=0.1$ eV, the vertical separation is $\omega_{ge}=1$ eV and the cavity mode frequency is resonant with the latter energy separation, namely $\omega_{\mathrm{c}}=\omega_{ge}$, and the chosen molecule-cavity coupling is $g=0.05$ eV.

The vibronic polariton energies in the energy range $1.0 < E < 1.8$ eV obtained after diagonalization are plotted against the Huang-Rhys factor $\lambda$ in Fig. \ref{fig:figure3}a.
From the plot, we learn that the vibronic polariton states $P_n^\pm$ ($n=0,1,2...$) are linear combinations of states $|1;g,n'\rangle$ and $|0; e, n''\rangle$ coming from the first photonic-electronic excitation manifold. 
Each pair of vibronic polariton states $\{ P^-_n, P^+_n \}$ for $n=0,1,2,3...$ shows the same Rabi splitting $2g$ at $\lambda=0$ but it varies smoothly at higher $\lambda$. 
Unlike the results in Ref.~\cite{Calvo2020}, given that we use two different but congruent harmonic frequencies $\omega_g= 2 \omega_e$, the assignment of state labels is unambiguous only at large $\lambda$, since the vibronic polariton states $P_n^{\pm}$ tend to uncoupled states with exact vibronic energies. 
An adiabatic $\lambda-$correlation in the form $P_n^- \to |0; e, n \rangle$ and $P_n^+ \to |1; g, n \rangle$ (sharing the same vibrational label $n$) is still possible for $n=0$ and 1, but a diabatic following (avoided crossings that turn into real crossings) with $\lambda$ indicates that states $P_n^{\pm}$ contain a combination of several uncoupled vibronic states $n'$. 
It should be noted that the vibronic states $| 0; g, n \rangle$ remain uncoupled by the cavity interaction (as happens in the Jaynes-Cummings model), and one of the polariton partners in the pair $\{ P^+_n, P^-_n \}$ tends at large $\lambda$ to the uncoupled vibronic state $| 1; g, n' \rangle$ that shows degeneracy with the vibronic state 
$| 0; g, n'+5 \rangle$ (see Fig. \ref{fig:figure1}a). 
Some other particularities of this simplified model are discussed in \cite{Calvo2020}.

In addition to different harmonic frequencies for the molecular PEC, we include other realistic features that are molecular non-adiabatic and dipole couplings which explicitly depend on the internuclear separation (here, it corresponds to the coordinate $q$). 
We have chosen a diabatic coupling term $V_{\mathrm{D}}(q)$ in Eq. \eqref{eq:hamiltonianMol} with $\alpha=0.21$ bohr and a dipole $d(q)$ in Eq. \eqref{eq:Cavityint}  with a width $\beta=0.95$ bohr. 
For simplicity, both couplings are centered at the crossing point $q_{\mathrm{c}}(r,\lambda)$, with $r=-0.5 \log 2$ 
(see Appendix \ref{sec:AppendixA}).
The new vibronic polariton states of this extended Holstein-quantum Rabi model show strong mixing with the variation of $\lambda$, so that they cannot even be roughly labeled by pairs $P^{\pm}_n$. 
In Fig.~\ref{fig:figure3}b, for $0< \lambda < 7$ we appreciate a plethora of avoided crossings among the polariton vibronic energies, and a close inspection indicates that many polaritons at $\lambda=0$ diabatically correlate to an asymptotic mixture at higher $\lambda$'s.

To disentangle diabatically the mixture resulting from many avoided crossings is beyond our scope, though it is doable by solving a set of coupled differential equations with the non-adiabatic couplings $\langle \Psi_m (\lambda) | \partial/\partial \lambda | \Psi_n (\lambda) \rangle$ among the polariton vibronic eigenstates at each $\lambda$. 
The picture of uncoupled states is again recovered at $\lambda \ge$ 7.5 (compare Fig. ~\ref{fig:figure3}a with Fig.~ \ref{fig:figure3}b) for vibronic states below 1.8 eV, as dipole and diabatic couplings become ineffective due to the vanishing overlap with vibronic wave functions. 
As a guide for the eye in Fig.~ \ref{fig:figure3}b, the energy at the crossing $V(q_{\mathrm{c}})$ for each value of $\lambda$ (black solid line), separates the inner region of strong coupling and the outer region of weak coupling. From the photodynamics perspective drawn in the previous section after a Franck-Condon excitation, the energies of the lowest LIC$_{\mathrm{R}}$ and of the diabatic crossing lie just above the available initial energy of 2 eV for $\lambda \ge 7.5$, leaving the system without effective couplings.  

The situation can be better understood as a $\lambda$-energy correlation that transitions from a large $\lambda$ (uncoupled) to a small $\lambda$ (fully coupled). Whereas the uncoupled states at large $\lambda$ $| 0;e;n \rangle$ are diabatically promoted to higher energies when $\lambda \to 0$, those uncoupled states $|1; g,n\rangle$ diabatically reduce their respective energies. Another very important issue to be considered is that those previously spectator states $| 0; g,n\rangle$ now become cavity-dressed indirectly through the diabatic couplings $\ket{g} \leftrightarrow \ket{e}$ without photon exchange. 
To conclude, we show that the explicit introduction of realistic non-adiabatic couplings $V_{\mathrm{D}}(q)$ dramatically changes the energetics of the molecular polaritonic system and therefore has important consequences in their photodynamics, as shown below.

\subsection{Model Hamiltonian vs. \textit{ab initio} alkaline earth monohydride molecules}
\label{sec:abinitio}

\begin{figure}[t]
    \centering   \includegraphics[width=0.95\linewidth]{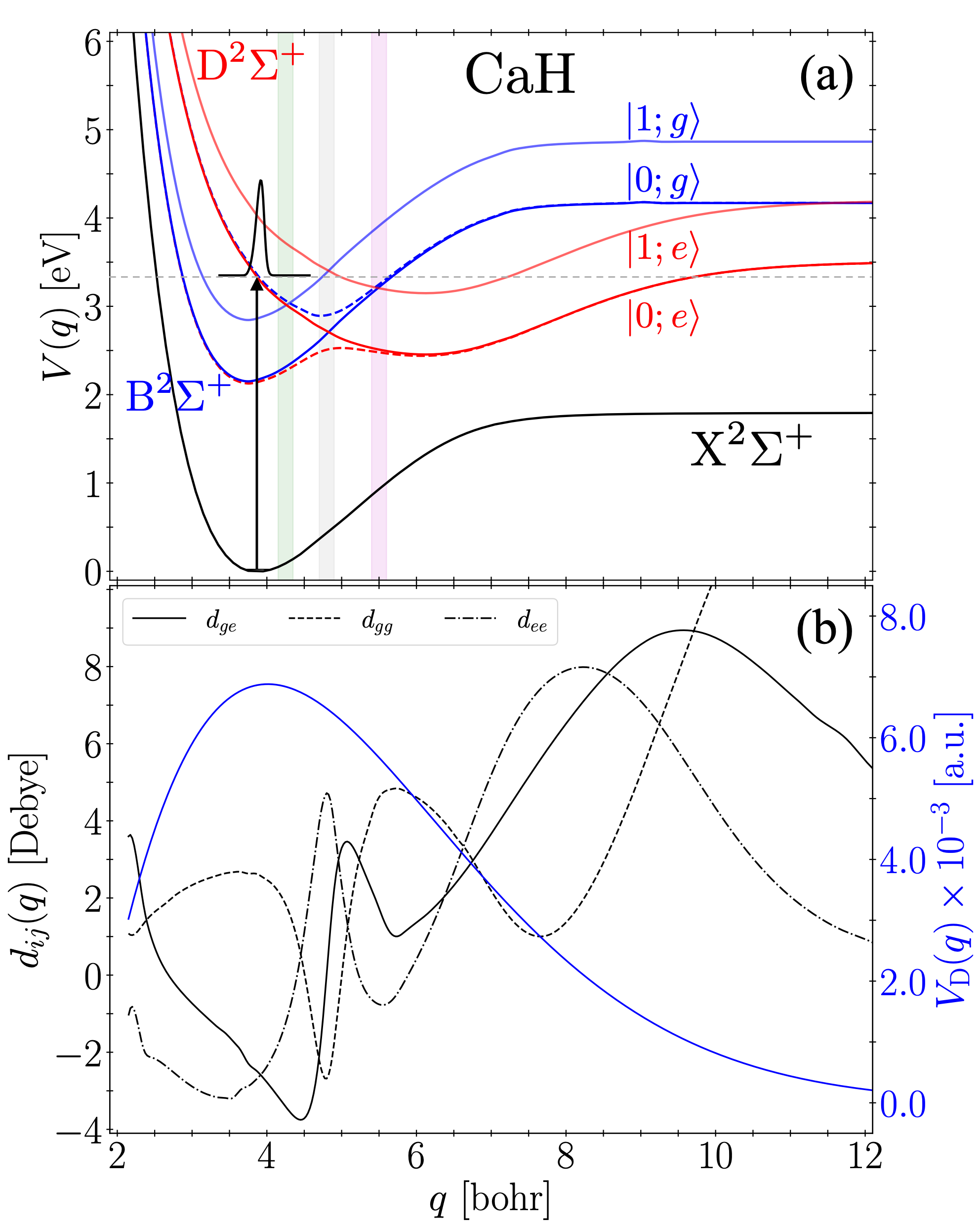}
    \caption{(a) Adiabatic (dashed lines) and diabatic (solid lines) potential energy curves of CaH molecule, corresponding to the ground state $\mathrm{X}^2\Sigma^+$, first excited state  $g \equiv 
    \mathrm{B}^2\Sigma^+$ and second excited state $e \equiv \mathrm{D}^2\Sigma^+$. A sudden vertical photo-absorption $|X,0\rangle \to \sum_n c_n |e,n \rangle$ by a pumping laser 
    generates an initial wave packet with average energy $E \sim 3.3$ eV. In this case, for a cavity mode frequency $\omega_{\mathrm{c}}$ = 5600 cm$^{-1}$= 0.694 eV only four dressed states 
    $|0;x\rangle$ and $|1;x\rangle$ for $x=g,e$ lie below the total available energy.
     (b) Diagonal dipole moments $d_{gg}(q)$ (black dashed line) and $d_{ee}(q)$ (black dot-dashed line) and transition dipole moments $d_{ge}(q)$ (black solid line) and diabatic electrostatic coupling $V_{\mathrm{D}}(q)$ (blue solid line), between excited states $g$ and $e$, Two $y$-axis scales are included for the dipole (left) and the diabatic coupling (right).}
    \label{fig:figure4}
\end{figure}

We validate our extended Holstein-quantum-Rabi model to study the physics involved in the fast polariton photodynamics of real diatomic molecules. 
We choose diatomic molecules XH (where X is an alkaline-earth atom, like Ca) because they show an energetically separated ground state X$^2\Sigma^+$ and the avoided crossing between the two lowest excited states B$^2\Sigma^+$ and D$^2\Sigma^+$ (see Fig. \ref{fig:figure4}a). The transition dipole moments and diabatic couplings between these two excited states show their prevalence close to the avoided crossing (see Fig. \ref{fig:figure4}b). 

A pumping laser transfers the state  $|X,0\rangle$ from the X$^2\Sigma^+$ state to a wave packet in the D$^2\Sigma^+$ excited state due to a Frank-Condon vertical transition (the eventual small portion excited to the B$^2\Sigma^+$ state plays no significant role in the photodynamics since it remains unaltered by the cavity interaction LIC or the natural DC).
The total available energy of the wave packet in D$^2\Sigma^+$ state is slightly below $3.5$ eV, which corresponds to the dissociation threshold energy.
Given the energy uncertainty of the initial wave packet, some small photodissociation yielding the fragments Ca$^*$ + H cannot be discarded.  In this respect, if we neglect dissociation, the PEC of CaH can be safely fitted to harmonic potentials, and then a Holstein-quantum-Rabi diabatic model can be applied.  

\begin{figure}[t]
    \centering
  \includegraphics[width=0.95\linewidth]{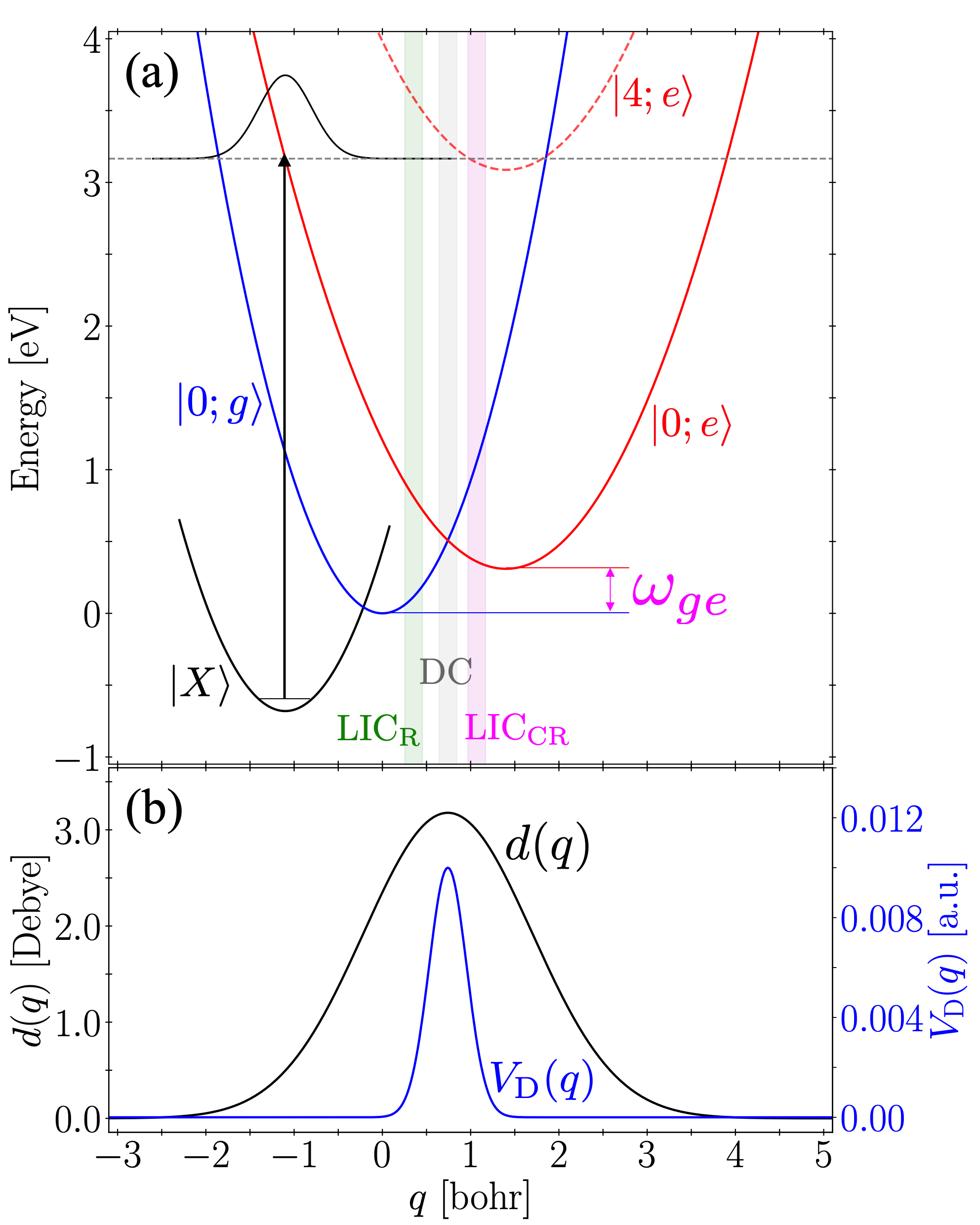}
    \caption{Molecular Holstein-Quantum-Rabi model fitted to the CaH molecule PECs shown in Fig. \ref{fig:figure4} with $\omega_\mathrm{g}=1350$ cm$^{-1}$, $\omega_\mathrm{e}=950$ cm$^{-1}$ and $\omega_{\mathrm{ge}}=2500$ cm$^{-1}$. 
    (a) A sudden vertical photo-absorption $|X,0 \rangle \to \sum_n c_n |e, n \rangle$ generates an initial wave packet with average energy $E \sim 3.3$ eV.
    The cavity interaction is switched on, and the wave packet moves rightward in the potential energy curve $V_e(q)$ to enter into a manifold of light induced and diabatic crossings, LIC$_{\mathrm{R}}$ (in a green shadowed area), DC (grey), LIC$_{\mathrm{CR}}$ (pink) (see section \ref{sec:holstein}).  For a cavity mode frequency $\omega_{\mathrm{c}}$ = 5600 cm$^{-1}$= 0.694 eV ten dressed states 
    $|n_c;x\rangle$ for $x=g,e$ and $n_c$=0,1,2,3 and 4 lie below the total available energy. Note that the dressed state $|4;e\rangle$ still enters slightly below the
    total available energy upon laser excitation.
    (b) Dipole $d(q)$ and diabatic coupling $V_{\mathrm{D}}(q)$ modelled with Gaussian functions centered at the diabatic crossing  $q_{\mathrm{c}}$.
    }  
    \label{fig:figure5}
\end{figure}

The PECs and dipole moments for the chosen diatomic molecule CaH were computed using the electronic structure package MOLPRO \cite{MOLPRO2012,MOLPRO}. 
We use the aug-cc-pwCVQZ-PP basis set for Ca and the aug-cc-pVQZ basis set for H, and the C$_{2v}$ point group was selected with the CaH molecule oriented along the $z$-axis, 
since MOLPRO can only use Abelian point groups. 
We also apply the LQUANT option in MOLPRO to force convergence to obtain accurate $^{2}\Sigma^{+}$ states \cite{MOLPRO}, due to the CaH molecule belonging to the C$_{\infty v}$ point group.
First, we perform a restricted Hartree-Fock (RHF) calculation to generate the molecular orbitals to be used as the initial guess for a complete active space (CAS) calculation of the fourth lowest $^{2}\Sigma^{+}$ states using a multi-configurational self-consistent field (MCSCF) method with 3 active electrons in 11 orbitals. 
Then, the multireference configuration interaction (MRCI) method, which includes single and double excitations, is implemented to obtain both energies, dipoles, and non-adiabatic couplings. 
The computed adiabatic PECs for the three states are drawn in Fig.~\ref{fig:figure4} along with the diabatized curves that show real crossings. 
The transition dipole moment between the excited states B$^2\Sigma^+$ and D$^2\Sigma^+$ $d_{ge}(q)$ shows extrema at three different internuclear distances. 
Since the dipole coupling is then integrated by the vibrational states of the B$^2\Sigma^+$ and D$^2\Sigma^+$ states, i.e., $\int \chi^B_{n'}(q)  d(q)  \chi^D_{n''} (q)\,\mathrm{d}q$, an effective overlap between the vibrational states of the two states is required. 
Consequently, only the extrema located at $q \sim 3-5$ bohr become relevant in dipole excitation, and these two peaks are very close to the diabatic crossing.
Similarly, the diabatic coupling $V_{\mathrm{D}}(q)$ shows a wide maximum that covers the diabatic crossing region. These preliminaries fit reasonably well within our proposed diabatic HQR model. 

Inspired by the CaH molecular scheme in Fig. ~\ref{fig:figure4}, the molecular parameters used in our extended HQR model calculations are $\omega_{\mathrm{g}}=1350$ cm$^{-1}=0.167$ eV, $\omega_{\mathrm{e}}=950$ cm$^{-1}=0.118$ eV, $\omega_{\mathrm{ge}}=2500$ cm$^{-1}=0.31$ eV for energetics, and equilibrium distances $q_g=0$, $q_e=1.4$ bohr and $q_{X}=-1.1$ bohr for PECs (see Fig.~\ref{fig:figure5}). 
The chosen location for $q_{X}$ requires an explanation because it is not aligned with $q_{g}$ as in CaH. We have shifted the $|X\rangle$ state to have a total available energy upon excitation ($E_0 \sim 3.3$ eV) larger than in CaH (1.3 eV), with respect to the bottom of the PEC associated with the $g$ state. 
This {\em unphysical} trick allows us to accommodate more dressed states (more photons) entering into the photodynamics as energetically open channels. Alternatively, this could be achieved with a pump-probe scheme, an IR laser preheating the initial vibrational state in the ground state $|X\rangle$ and a subsequent VUV laser pulse delayed just when the IR-excited wave packet localizes on the left turning point in the X$^2\Sigma^+$ PEC. 

The HQR model does not contain the anharmonicity already present in the \textit{ab initio} PECs of the CaH molecule. The number of dressed states that energetically enter below the total available energy $E_0$ is different for each cavity mode frequency $\omega_{\mathrm{c}}$. 
In this case, for a chosen $\omega_{\mathrm{c}}= 5600$ cm$^{-1}= 0.694$ eV only four dressed states $\{ |n_{\mathrm{c}}; g\rangle, |n_{\mathrm{c}}; e\rangle \}$ (with $n_{\mathrm{c}}=0, 1$) enter below the total energy 3.3 eV using the CaH \textit{ab initio} PECs (note that the zero energy is set at the bottom of the ground state $|X\rangle$).
On the other hand, in the HQR model adjusted to CaH the total energy is 3.1 eV from $V_g(q_g)=0$, and for the same cavity model frequency $\omega_c$, up to ten dressed states $\{ |n_{\mathrm{c}}; g\rangle, |n_{\mathrm{c}}; e\rangle \}$ (with $n_{\mathrm{c}}=0,1,2,3$ and 4) fit below. Our trick helps to better understand the contribution of a larger number of photons.

The diabatic crossing of the extended HQR model in Fig.~\ref{fig:figure5} is located at $q_{\mathrm{c}}=0.74$ bohr. The transition dipole moment $d(q)$ and diabatic coupling $V_{\mathrm{D}}(q)$ are modeled by Gaussian distributions centered on $q_{\mathrm{c}}$ with widths $\sigma_{\mathrm{d}}=0.95$ bohr and  $\sigma_\mathrm{D} = 0.21$ bohr, respectively.
Their strenghts are adjusted to closely approximate CaH values.
We use a reduced nuclear mass $M = M_{\mathrm{Ca}} M_{\mathrm{H}} / (M_{\mathrm{Ca}} + M_{\mathrm{H}})$ with $M_{\mathrm{Ca}}$ = 40.078 amu
and  $M_{\mathrm{H}}$=1.00784 amu so that the diabatic crossing has energy $V_{\mathrm{c}}=\frac{1}{2} M \omega_g^2 q^2_{\mathrm{c}}$. 
Our molecular HQR model, partially adjusted for CaH, is represented by the PECs and couplings shown in Fig. \ref{fig:figure5}.

\subsection{Computational methods for photodynamics}
\label{sec:computational}

We have implemented two computational methods of solution. 
First, we implemented a Python code using the QuTiP libraries for quantum optics \cite{qutip}, which makes use of explicit uncoupled basis $|n_{\mathrm{c}}; g,n\rangle$ and $|n_{\mathrm{c}}; e,n \rangle$, to diagonalize the Hamiltonian and to solve the time-dependent Schrödinger equation using the spectral method in terms of the same basis set. 
In this case, operators for vibrational and photonic modes in the Hamiltonian (including position operators $\hat{q}$) are expressed in terms of bosonic operators and the computation of all matrix elements is straightforward. 
For the sake of convergence, we employ up to $n=110$ vibrational states for each state $g$ and $e$, up to $n_{\mathrm{c}}=11$ Fock states for the photon space, which amounts to having Hamiltonian matrices with dimension $2420 \times 2420$ to be diagonalized, and 2420 time-dependent coupled differential equations to be time integrated.
This kind of computation using QuTiP can be easily performed on a desktop computer.   

Another proposal for polariton photodynamics is to solve the time-dependent Schr\"odinger equation in coordinate space with the light-matter Hamiltonian given in Eq. \eqref{eq:hamiltonian} by implementing the multi-configuration time-dependent Hartree (MCTDH) method \cite{Beck2000,MCTDH2021}. 
The multistate MCTDH wave function ansatz for the molecule coordinates $q$ and the photon coordinates $x$, respectively, can be written as
\begin{equation}
\Psi(q,x,t)=\sum_{\alpha=1}^{2}\sum_{j_{q}=1}^{n_{q}}\sum_{j_{x}=1}^{n_{x}} A_{j_{q}j_{x}}(t) \phi_{j_{q}}^{(\alpha)}(q,t)\phi_{j_{x}}^{(\alpha)}(x,t), 
\end{equation}
which is an expansion constructed by Hartree products of the time-dependent basis functions $\phi_{j_s}^{(\alpha)}(s,t)$ of the electronic state $\alpha$, with $s=\{q,x\}$.
We represent the molecular coordinate on a uniform grid with $N_{q}=451$ points on the interval $-6.0 < q < 9.0$ bohr. The photonic mode is represented by the coordinate $x$ in a uniform grid with $N_{x}=181$ points along the dimensionless interval $-60.0 < x < 60.0$. 
The number of time-dependent basis functions $n_{q}=n_{x}$ depends on the cavity frequency and is selected in the interval $15<n_{s}<30$. 
We have verified that both computational methods provide the same solution upon convergence for a set of different coupling factors $V$ and $\chi$ in Eqs. \eqref{eq:hamiltonianMol} and \eqref{eq:Cavityint}, respectively.

\section{Results}
\label{sec:results}


In this section, we include a series of numerical computations for the polariton photodynamics \textit{i)} using the HQR Hamiltonian model \textit{ii)} using the \textit{ab initio} Hamiltonian for the molecule CaH. Our purpose is to verify how Hamiltonian models allow us to reproduce and understand the physical mechanisms expected in real complex molecular systems.

\subsection{Molecular polariton photodynamics with the HQR model}

We first study the non-adiabatic photodynamics of a single diatomic molecule coupled to a quantized cavity using the extended HQR model for a chosen cavity mode frequency $\omega_{\mathrm{c}}=5600$ cm$^{-1}$ described in Sec. \ref{sec:holstein} and Fig. \ref{fig:figure5}. 
This frequency value is chosen to allow dressed electronic states with up to four photons, but also because it produces a noticeable separation between the regions where the diabatic crossing (DC) and light-induced crossings (LIC$_\mathrm{R}$ and LIC$_\mathrm{CR}$) act mostly, as shown in Fig.~\ref{fig:figure5}a.  
Note that for $\omega_{\mathrm{c}}\to 0$, the light-induced crossings LIC$_{\mathrm{R}}$ and LIC$_{\mathrm{CR}}$ approach the position of the diabatic crossing DC, and their simultaneous effect becomes intertwined.

\begin{figure}[t]
    \centering  \includegraphics[width=0.95\linewidth]{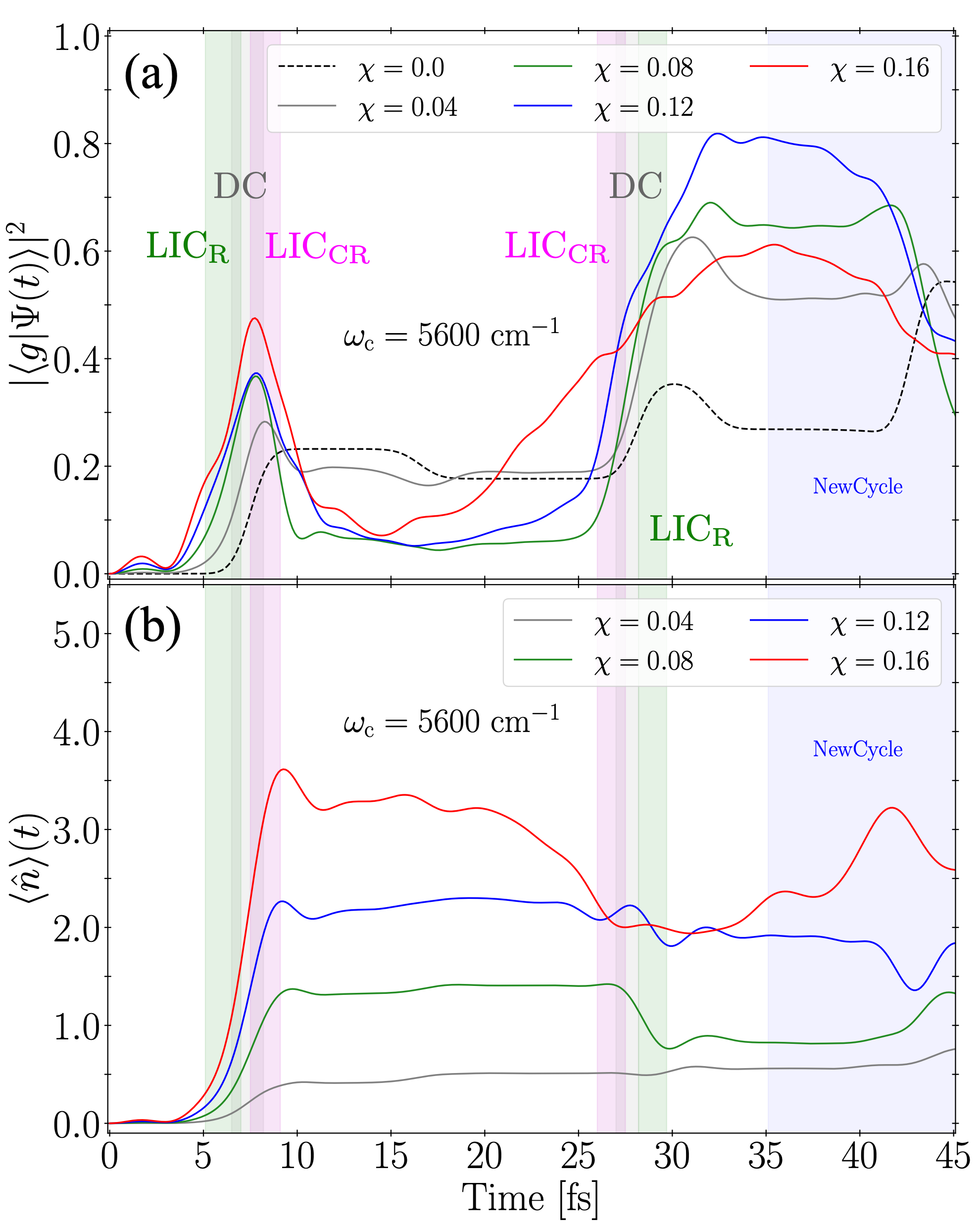}
    \caption{Polariton photodynamics of the extended Holstein-Quantum-Rabi model for the molecular parameters
    quoted in Fig. \ref{fig:figure5}. The chosen cavity frequency mode is $\omega_{\mathrm{c}}=5600$ cm$^{-1}=0.694$ eV
    and without a cavity interaction, $\chi=0$ (dashed line), and for
    different molecule-cavity couplings (solid lines)  $\chi=0.04$ (black), $\chi=0.08$ (green), $\chi=0.12$ (blue) and $\chi=0.16$ (red).
    (a) Time-dependent evolution of the population of the $|g\rangle$ state, i.e., $|\langle g | \Psi(t) \rangle|^2$, for all energetically available photons
    and vibrational states. 
    (b) Time-dependent evolution of the average number of photons $\langle  \hat{n}(t) \rangle$ present within the cavity. Vertical shadowed areas
    indicate zones of influence for LIC$_\mathrm{R}$ (RWA, in green), DC (grey) and LIC$_{\mathrm{CR}}$ (non-RWA, in pink). After 35 fs the wave packet returns to the initial position and a new cycle begins.}
    \label{fig:figure6}
\end{figure}

The initial state is prepared, prior to cavity interaction, as a direct product of the ground vibrational state of the electronic state $\ket{X}$ along with the Fock vacuum state $\ket{n=0}$, to start its dynamics in the excited electronic $\ket{e}$ at $t=0$, as shown in Fig.~\ref{fig:figure5}.
By solving the time-dependent Schrödinger equation, we obtain the total polaritonic wave packet in coordinate space $\Psi(q,x,t)$ $\forall t > 0$. 
Figure~\ref{fig:figure6}a shows the time-dependent population transfer to the ground state $\ket{g}$, namely, $|\langle g | \Psi (t)\rangle|^2$, for different light-matter coupling strengths $\chi$. This means that we sum over all photons and vibrational states.  

For $t<15$ fs, the wave packet consecutively crosses the LIC$_{\mathrm{R}}$, the diabatic crossing DC, and the counter-rotating light-induced crossing LIC$_{\mathrm{CR}}$. 
To compare the polariton photodynamics with the usual molecular dynamics (without cavity field), we also include the temporal ground state population in the latter case for $\chi=0$ in Fig.~\ref{fig:figure6}a.
Keep in mind that the harmonic period of the $|g\rangle$ state is $\sim 7/10$ of the $|e\rangle$ state, and in our simulations we obtain approximately $T_g \sim 24$ fs and $T_e \sim 35$ fs.
Without field, there are two main features: 
\begin{itemize}
\item The wave packet splits into two components and transfers $\sim20$\% to the $|g\rangle$ state before 10 fs. The $g$-wave packet propagating in the PEC $V_g(q)$ reaches the right turning point located at $\sim 2$ bohr, then moves leftward to transfer another $\sim20$\% of its own population back to the initial state $|e\rangle$ before 20 fs. 
\item The residual $e$-wave packet moving in state $|e\rangle$ carrying a population of around 80\% reaches its turning point at $q \sim 4$ bohr later in time and moves leftward to find the DC again, thus producing another transfer to $|g\rangle$ state (just before 30 fs) plus another subsequent loss of population in $|g\rangle$ state due to another $|g\rangle \to |e\rangle$ transfer from the $g$-wave packet moving rightward in the PEC $V_g(q)$, after reaching the left turning point at $q = -2$ bohr. The latter combined effect is due to the squeezing between the two harmonic potentials, that leads to a reduced period in the $g$-wave packet.
\end{itemize}

\begin{figure*}
    \centering \includegraphics[width=0.9\linewidth]{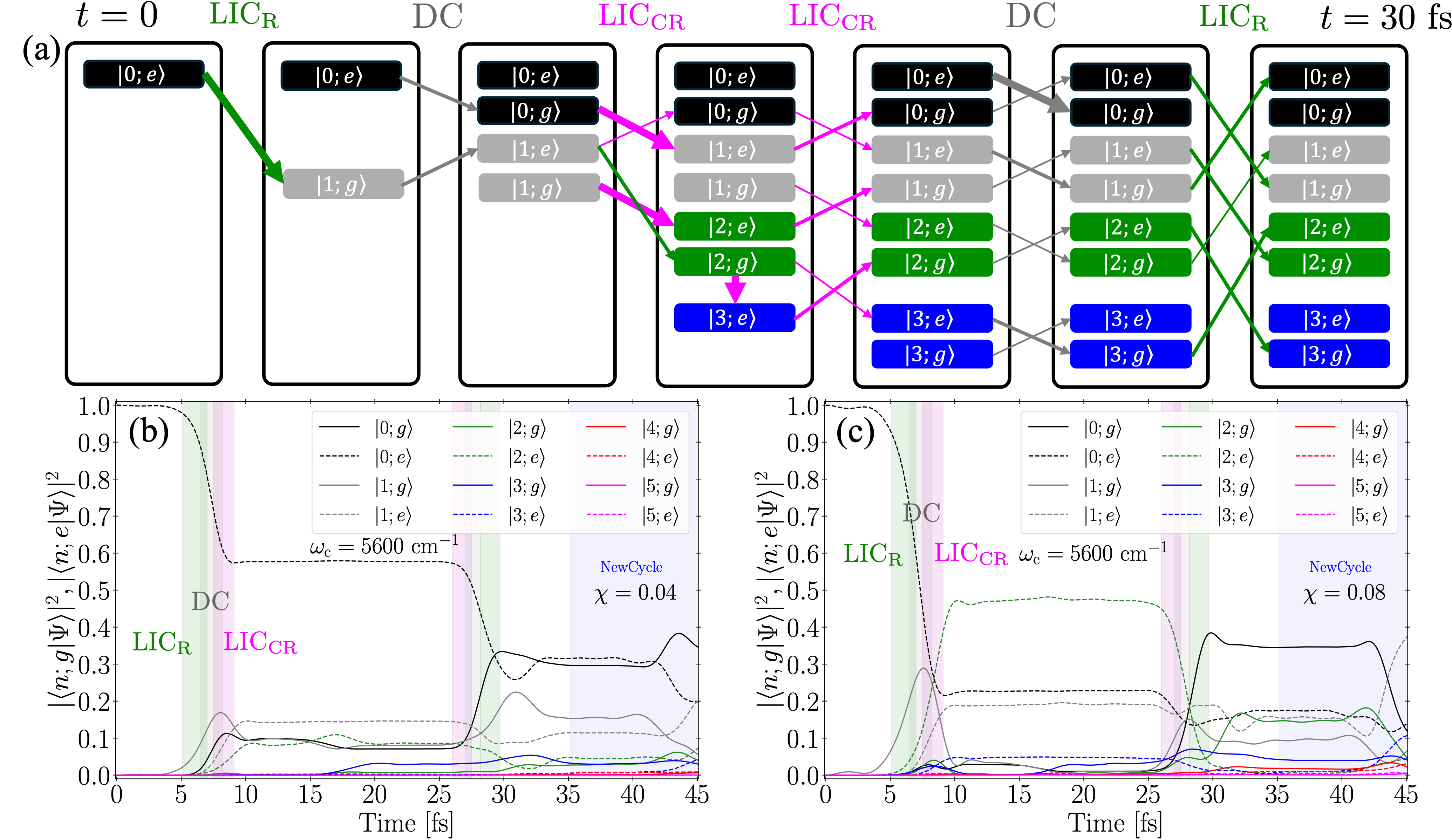}
    \caption{ Polariton ultrashort photodynamics of the extended Hosltein-Quantum-Rabi model for the molecular parameters
    quoted in Fig. ~\ref{fig:figure5}. The chosen cavity frequency mode is $\omega_{\mathrm{c}}=5600$ cm$^{-1}=0.694$ eV.
    The time window for the evolution includes a full periodic cycle $T_{\mathrm{c}} \sim 35$ fs of the initial wave-packet. 
    (a) Scheme of mechanisms of the polariton photodynamics:  transitions among the manifold of uncoupled states $( |n_{\mathrm{c}};g\rangle, |n_{\mathrm{c}};e \rangle$) along the track followed by the molecular wave packet moving rightward (LIC$_{\mathrm{R}}$, DC, LIC$_{\mathrm{CR}}$) and leftward  (LIC$_{\mathrm{CR}}$, DC, LIC$_{\mathrm{R}}$). For the chosen cavity mode frequency,  zero (black), one (grey), two (green) and three (blue) photon states are energetically reached. Green arrows indicate rotating transitions at LIC$_{\mathrm{R}}$, pink arrows indicate counter-rotating transitions at LIC$_{\mathrm{CR}}$, and grey lines correspond to pure matter transitions due to the diabatic crossing at DC.
    (b) Time evolution of projected populations of the polariton wave packet $| \langle n_{\mathrm{c}};g | \Psi (t)\rangle|^2$ (solid lines) and   $| \langle n_{\mathrm{c}};e | \Psi (t)\rangle|^2$ (dashed lines) for
    $n_{\mathrm{c}}=0$ (black), $n_{\mathrm{c}}=1$ (grey), $n_{\mathrm{c}}=2$ (green), $n_{\mathrm{c}}=3$ (blue), $n_{\mathrm{c}}=4$ (red) and $n_{\mathrm{c}}=5$ (purple), for a weak cavity coupling $\chi=0.04$. 
    (c) Same as (b) for cavity coupling $\chi=0.08$. }
    \label{fig:figure7}
\end{figure*}

Now, the molecule-cavity interaction is switched on. The population transfer $\ket{e}\to\ket{g}$ increases noticeably for $t < 8$ fs as the coupling strength $\chi$ increases, due to the action of the LIC$_{\mathrm{R}}$ (see Fig. ~\ref{fig:figure5} and Fig.~\ref{fig:figure6}). 
However, once the wave packet subsequently reaches the LIC$_{\mathrm{CR}}$ position ($q \sim 1.1$ bohr at $t > 8$ fs), the $\ket{g}$-state population mostly returns to the $\ket{e}$-state via a counter-rotating transition, e.g., $\ket{n_{\mathrm{c}};g}\to\ket{n_{\mathrm{c}}+1;e}$. 
For the time interval $15<t<25$ fs, the population transfer $\ket{g}\to\ket{e}$ occurs mainly due to DC crossing, which is explicitly observed in the absence of the cavity field with the transfer around $t\sim 16$ fs. 
Finally, the last part of the dynamics within the first full cycle $T_e = 35$ fs occurs between 25 fs and 35 fs. 
Here, the traveling $e$-wave packet that bounces back after the right turning point in the PCE $V_e(q)$ reaches the region of non-adiabatic crossing again, but in reverse order. 
We appreciate a huge population transfer $\ket{e} \to \ket{g}$ via LIC$_{\mathrm{CR}}$ aided by the DC coupling, and subsequently via the LIC$_{\mathrm{R}}$. For all interaction strengths $\chi$, the transfer $\ket{e}\to\ket{g}$ is much larger than that due only to DC, before starting a new cycle.
Fig.~\ref{fig:figure6}b shows the time-dependent mean photon number $\langle\hat{n}\rangle(t)$ present in the photonic mode, for different light-matter coupling strengths, and provides complementary information for the population transfer. Note that $\langle\hat{n}\rangle (t)$ increases with the coupling strength $\chi$ during the time interval of the first cycle. 
The maximum of photons is obtained once the $e$-wave packet crosses the first two LIC positions for $t < 10$ fs. The DC does not modify the number of photons at each crossing, and its effect is hardly seen in $\langle\hat{n}\rangle (t)$. The maximum value reached dynamically for $\langle\hat{n}\rangle$ depends on the initial state energy, which in this case has a value of $\sim 25000$ cm$^{-1}$, and for a cavity mode with $\omega_{\mathrm{c}}=5600$ cm$^{-1}$, it yields $\langle\hat{n}\rangle_{\mathrm{max}}< 4.5$, in good agreement with our results at the time interval $t=9-10$ fs.

We highlight that the population transfer to $\ket{e}\to\ket{g}$ for $\chi=0.16$ follows a different mechanism than for lower values (the sequence $\chi=0.04$, $0.08$, and $0.12$ seems to follow a monotonic behavior to reach a saturation limit). 
In this case, a larger number of photons are generated (up to $n_{\mathrm{c}}=4$), which means that dressed states $\ket{n_{\mathrm{c}};g}$ and $\ket{n_{\mathrm{c}};e}$ are energetically open channels, and for the initial state energy, $E_0 \sim 3.1$ eV, those dressed states display different turning points (see, for instance, Fig. \ref{fig:figure2}), and the analysis of the polariton wave packet becomes more complicated. 
We will return to this below.

\subsection{Non-adiabatic mechanisms in molecular polaritons}

To understand the inner workings of the different non-adiabatic crossings (radiative or non-radiative), we illustrate the mechanism of sequential transitions between the uncoupled states $\ket{n_{\mathrm{c}};g}$ and $\ket{n_{\mathrm{c}};e}$ in Fig. \ref{fig:figure7}a, which explains the underlying dynamics of Figs. \ref{fig:figure7}b and \ref{fig:figure7}c for the lowest two chosen couplings $\chi=0.04$ and $\chi=0.08$, respectively. 

The initial state is the uncoupled state $\ket{0;e}$ (leftmost frame in Fig. ~\ref{fig:figure7}a). Then the total polariton wave packet first reaches the LIC$_{\mathrm{R}}$ position for $t > 5$ fs, and it splits into two components, $\ket{0;e}$ and $\ket{1;g}$ (second frame from left to right in Fig. ~\ref{fig:figure7}a). Next, these two components cross the DC position performing pure matter transitions $\ket{0;e}\to\ket{0;g}$ and $\ket{1;g}\to\ket{1;e}$, without photon exchange (third frame in Fig. \ref{fig:figure7}a).
Afterwards, these four components within the third frame arrive at the LIC$_{\mathrm{CR}}$ position, with associated CR transitions $\ket{0;g}\to\ket{1;e}$, $\ket{1;e}\to\ket{0;g}$ and $\ket{1;g}\to\ket{2;e}$.
In addition, because of the spreading of the wave packet and the width of the transition dipole moment $d_{ge} (q)$, the tail of the wave packet overlaps at the same time the LIC$_{\mathrm{R}}$ position, thus also transferring some population from state $\ket{1;e}$ to state $\ket{2;g}$ (denoted with a green arrow in Fig. \ref{fig:figure7}a between the third and fourth frames). 
In particular, this previous R transition explains the CR transition $\ket{2;g} \to \ket{3;e}$ that appears in Fig. ~\ref{fig:figure7}c for $\chi = 0.08$, which is illustrated by the pink vertical thick arrow in the fourth frame in Fig. ~\ref{fig:figure7}a.

At this point, after the LIC$_{\mathrm{CR}}$ position, for $t> 10$ fs the states contributing with higher photon components $\ket{1;e}$ and $\ket{2;g}$ are created by the LICs and aided by the DC, by explicitly creating the intermediate $\ket{0;g}$ state.
So far, we have described the first twelve femtoseconds of the dynamics. 
From $t=12$ fs to $t=25$ fs, a returning $g-$wave packet first reaches the triplet in reverse order (LIC$_{\mathrm{CR}}$, DC,  LIC$_{\mathrm{R}}$) and we have the noticeable feature of the raising of state $\ket{3;g}$, through the staircase mechanism $\ket{1;g} \xrightarrow{\mathrm{CR}} \ket{2;e} \xrightarrow{\mathrm{R}} \ket{3;g}$, seen in Fig.~\ref{fig:figure7}b at $t=15-20$ fs. 
This mechanism is indicated in Fig. ~\ref{fig:figure7}a by the sudden creation of state $\ket{3,g}$ in the fifth frame, which represents the start of the reverse order. For the higher coupling $\chi=0.08$ the LICs dominate over the DC, and bring the dominance of dressed states $\ket{1;e}$,  $\ket{2;e}$ and $\ket{3;e}$ that carry the photons. 
From $t=25$ fs to $t=32$ fs, and $\chi=0.04$, the $e$-wave packet crosses again the non-adiabatic region but in reverse order. The dominant component $\ket{0;e}$ is not affected by any CR transition, so that the DC strongly acts to transfer to $\ket{0;g}$, 
and the LIC$_{\mathrm{R}}$ produces the transfer $\ket{0;e} \to \ket{1;g}$.
For $\chi=0.08$ the higher strength and the extension of the transition dipole moment together with the wave packet spreading preclude the unambiguous assignment of the sequential action of couplings. The dominant components before the crossing in reverse at $t > 25$ fs in the $e$-wave packet are $|\braket{2;e}{\Psi}|^2 > |\braket{0;e}{\Psi}|^2 > |\braket{1;e}{\Psi}|^2$. 
From a closer analysis of Fig. ~\ref{fig:figure7}, we find the following mechanisms (among others less important), namely, $\ket{2;e} \xrightarrow{\mathrm{CR}} \ket{1;g} \xrightarrow{\mathrm{DC}}\ket{0;g}$ leading to a dominant state $\ket{0;g}$ in the new cycle, and $\ket{0;e} \xrightarrow{\mathrm{R}} \ket{1;g} \xrightarrow{\mathrm{DC}} \ket{1;e} \xrightarrow{\mathrm{R}} \ket{2;g}$, which promotes a state with 2 photons in the new cycle.

It is clear that the polariton photodynamics cannot be reversible (the same couplings crossed twice in reverse order would naively bring the wave packet back to the initial state). Instead, there are many other channels open by R, CR, and DC couplings that interfere, and this generates a complex wave packet after a few cycles, which is difficult to disentangle or simplify.

\begin{figure}[h!]
    \centering
    \includegraphics[width=0.95\linewidth]{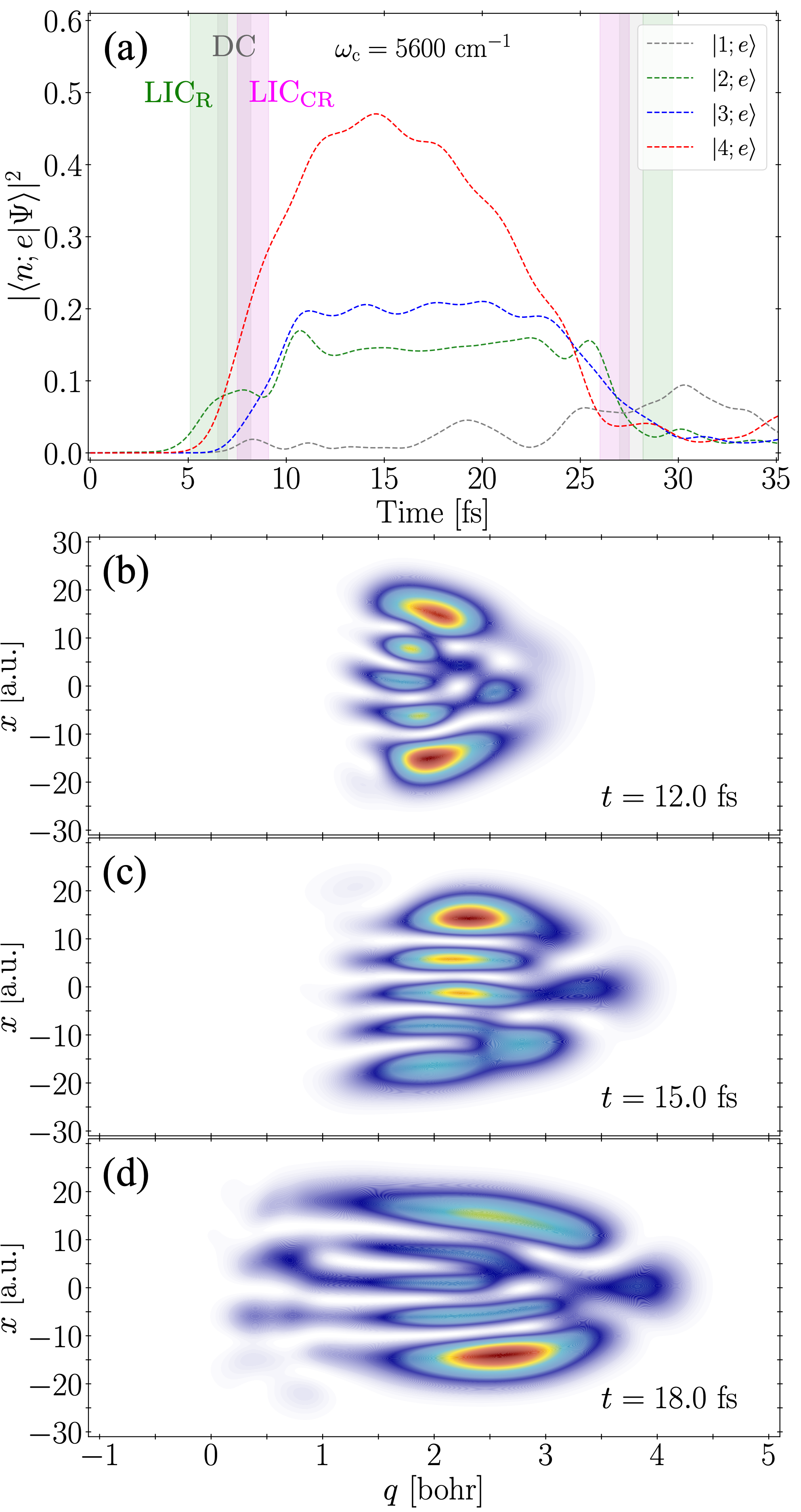}
    \caption{(a) Excited state probabilities of the time dependent polariton state, $|\langle n_{\mathrm{c}}; e | \Psi (t) \rangle |^2$ (for the number of photons $n_{\mathrm{c}}$ = 1,2,3 and 4), of  the extended Hosltein-Quantum-Rabi model for the molecular parameters quoted in Fig. \ref{fig:figure5}. 
    The chosen cavity frequency mode is $\omega_{\mathrm{c}}=5600$ cm$^{-1}=0.694$ eV and the molecule-cavity coupling is $\chi=0.16$. The time window in the figure corresponds to a complete cycle of the initial wave packet. Snapshots of the time-dependent probability density  of the polariton wave packet in the excited state $\ket{e}$, i.e.,  $|\Psi_e (q,x)|^2$ extracted from MCTDH, 
    represented against the vibrational coordinate $q$ and the photon coordinate $x$, taken at different times $t= 12$ (b), 15 (c) and 18 fs (d). Note that the wave packet center barely moves from $q = 2$ bohr and the four photon state dominates. }
    \label{fig:figure8}
\end{figure}

\subsection{Temporal photon trapping}
\label{sec:trapping}

We discuss here a particular feature that appears for the coupling strength $\chi=0.16$. As noted above in Fig.~\ref{fig:figure6}, the polariton photodynamics follows a different trend after the $e$-wave packet moves across the first set of non-adiabatic couplings for $t > 20$ fs. The transfer $\ket{e} \to \ket{g}$ increases linearly, whereas the average number of photons drops.
The value $\chi=0.16$ belongs to a regime of ultra-strong coupling where the CR non-RWA transitions dominate. 
In Fig.~\ref{fig:figure8}a, we plot the temporal probabilities for the excited state $|e\rangle$ in the total polariton WF, $|\langle n_{\mathrm{c}}; e | \Psi (t) \rangle |^2$, for the number of photons $n_{\mathrm{c}}=1,2,3$ and 4. The initial state contains the vacuum $\ket{0;e}$. 
The dominant contribution is $\ket{4;e}$ after the first crossing through the non-adiabatic zone. 
In order to reach the state $\ket{4;e}$, a simultaneous action of the three couplings (LIC$_{\mathrm{CR}}$, DC,  LIC$_{\mathrm{R}}$) is required, and several times within the same time window. Different paths are possible, among them, namely, 
$\ket{0;e} 
\xrightarrow{\mathrm{R}} \ket{1;g} 
\xrightarrow{\mathrm{DC}} \ket{1;e} 
\xrightarrow{\mathrm{R}} \ket{2;g} \xrightarrow{\mathrm{CR}} \ket{3;e} \xrightarrow{\mathrm{DC}} \ket{3;g} \xrightarrow{\mathrm{CR}} \ket{4;e}$.
In this particular case, we refer to Fig.~\ref{fig:figure5}, where the PEC for the dressed component $\ket{4;e}$ lies energetically at the edge of the total initial energy. Even taking into account the energy uncertainty of the wave packet, the turning points of the dressed PEC $\ket{4;e}$ are much more constrained in $q-$space than those for the PEC of the vacuum state $\ket{0;e}$. 
This implies that the vacuum is suddenly promoted to be dressed mostly by 4 photons, and this polariton WF becomes temporarily trapped around $q \sim 2$ bohr. This feature is clearly illustrated in Figs.~\ref{fig:figure8}b, \ref{fig:figure8}c and \ref{fig:figure8}d. 
Just after the dominant promotion to state $\ket{4;e}$ at $t=12$ fs, the polariton WF appears well localized in matter $q-$space and also displays 4 nodes along the light $x-$coordinate. This component $\ket{4;e}$ always remains stuck in the zone of non-adiabatic couplings, so that it drains population (and photons!) through the mechanism in reverse, towards other dressed states below, to end at $t=35$ fs with
a 60\% in dressed states $\ket{n_{\mathrm{c}};g}$ and 40\% in
states $\ket{n_{\mathrm{c}};g}$. This photon-trapping effect depends on the strength and $q$-extension of non-adiabatic couplings as well as the choice of the Huang-Rhys factor to find conditions under which the draining of photons be reduced after sudden excitation.
This effect could be exploited to manipulate the temporal storage of photons in molecular systems within cavities. It remains an open question how these photons, inherited from a molecular excess energy, and created within a short time window could be characterized with some entanglement properties. 
This photon-trapping in polaritonics finds its counterpart vibrational-trapping in previous studies of photodissociation of simple molecules by intense laser fields
\cite{Bucksbaum1990}. This effect called bond-hardening (creation of a temporal molecular bond by a laser) in the photodissociation of H$^+_2$ was studied by applying time-dependent non-Hermitian Floquet techniques \cite{Chu2004}.

\subsection{Vibronic energy conversion into photons}

\begin{figure}[h]
    \centering   \includegraphics[width=0.95\linewidth]{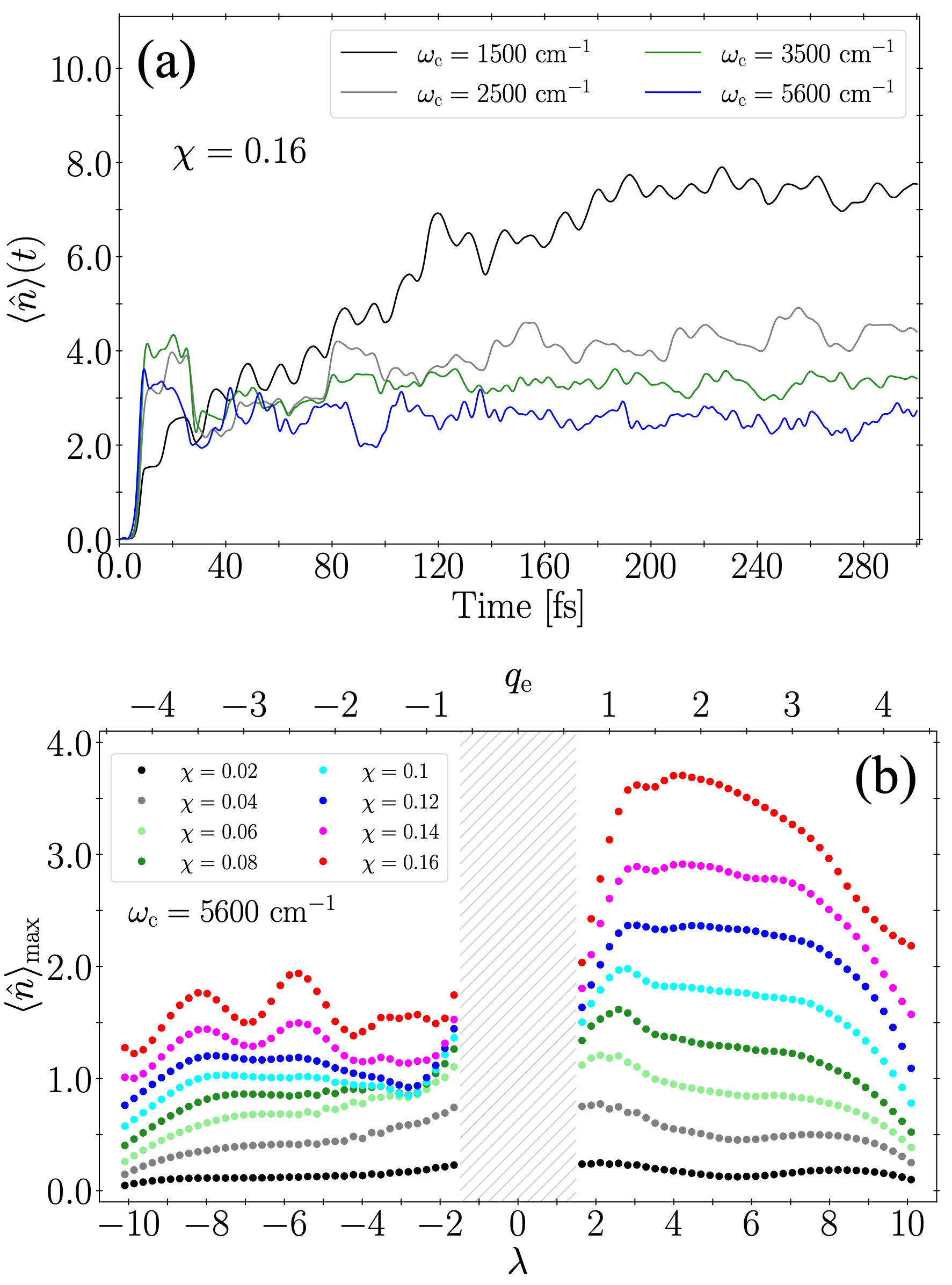}
    \caption{(a) Time evolution of the average number of photons contained in the polariton wave packet resulting from non-adiabatic conversion of an initial state with energy $E_0 = 3.2$ eV (see Fig. \ref{fig:figure5}), for a molecule-cavity coupling $\chi=0.16$, and for a set of cavity mode frequencies $\omega_{\mathrm{c}}$ = 1500, 2500, 3500, 5600 cm$^{-1}$.
    (b) Maximum of the average number of photons contained in the polariton wavepacket  as a function of Huang-Rhys factor $\lambda$ (bottom axis) or, equivalently, against the 
    position of equilibrium of the excited state $q_{\mathrm{e}}$ (top axis) for different coupling strengths $\chi$ and a cavity mode frequency $\omega_{\mathrm{c}}=5600$ cm$^{-1}$.
    The shadowed area corresponds to small magnitudes of the Huang-Rhys parameter $| \lambda |$  for which the potential energy curves cross twice below the initial state energy $E_{0}$.} 
    \label{fig:figure9}
\end{figure}

Until now, we have considered the ultrafast energy conversion into photons within the first few tens of fs (within the first wave packet cycle). The polariton WF at this point is normally a large superposition of dressed components $\ket{n_{\mathrm{c}};g}$ and $\ket{n_{\mathrm{c}};e}$, and a detailed description of the mechanisms in the subsequent photodynamics is unattainable. 
In addition, beyond 100 fs, photon loss and other relaxation mechanisms become relevant enough to consider the open quantum system and to solve, e.g., the quantum Lindblad master equation \cite{Manzano2020} or the hierarchical equations of motion (HEOM) \cite{Tanimura1989,Tanimura1990}.
However, here we show whether photon conversion exhibits long-term stability and what determines the upper limit for the number of photons generated from the vacuum. 
In Fig.~\ref{fig:figure9}a, we include the average number of photons created by non-adiabatic energy conversion for the most favorable case of high coupling strength $\chi=0.16$ and for different cavity mode frequencies,
$\omega_{\mathrm{c}} = 5600$, 3500, 2500, and 1500 cm$^{-1}$. This assortment of frequencies allows an increasing number of dressed states to fit energetically below the initial total energy $E_0$; the ratio $E_0/\omega_{\mathrm{c}}$ yields an approximate number of photons 
4, 7, 10, and 17, respectively. Simulations in Fig.~\ref{fig:figure9}a show that this nominal maximum number of photons is never achieved at long times ($t \sim 300$ fs), but at least it saturates toward half of that number. 
The photonic fluctuations after the first cycle are a signature of continuous photon exchange within the total polariton WF due to non-adiabatic couplings among the dressed states in the expansion. 
However, once the total WF reaches the effective maximum number of dressed states in the expansion, the photon exchange enters a quasi-stationary regime, with small out-of-equilibrium fluctuations. The case with $\omega_\mathrm{c}=1500$ cm$^{-1}$ in Fig.~\ref{fig:figure9}a seems to follow a different trend. The production of photons during the first cycle is less than with higher cavity frequencies, then it increases steadily during the six following cycles, and the photonic saturation starts from 200 fs. 
As $\omega_\mathrm{c} \to 0$ the positions of both LIC$_{\mathrm{R}}$ and LIC$_{\mathrm{CR}}$ colapse with that of the DC, so that the three couplings act simultaneously over the state, which eventually means merging three frames into one in Fig.~\ref{fig:figure7}a.
Since we concluded that a better photon yield is obtained with an ordered sequence R $\to$ CR and not CR$\to$R, it may explain the reduced production during the first cycle, but the larger available energy excess with small $\omega_\mathrm{c}$ compensates for subsequent cycles. 

To conclude this section, we emphasize the asymmetry in the photon promotion from the vacuum with respect to the Huang-Rhys factor $\lambda$.  For simplicity, we assume that the total initial energy $E_0$ remains the same when the Huang-Rhys factor changes (as mentioned above). In practice, we force the initial $e$-wave packet to always start from the classic left turning point and then moving rightward.

With this setup and geometry, the time evolution of the polariton wave packet is different for positive and negative Huang-Rhys factors. In all previous simulations, we chose a positive Huang-Rhys factor for which the ordering of couplings in the non-adiabatic zone was (LIC$_{\mathrm{R}}$, DC, LIC$_{\mathrm{CR}}$). In the case of a negative Huang-Rhys factor,  the situation is opposite, the initial $e$-wave packet first meets the non-adiabatic zone in reverse order (LIC$_{\mathrm{CR}}$, DC, LIC$_{\mathrm{R}}$), namely, the CR excitations act before the R ones. Since the initial state $\ket{0;e}$ is vacuum-dressed, the CR terms cannot affect the state, and the first effective couplings are DC ($\to \ket{0;g}$) and R
($\to \ket{1,g}$). In this case the loss of the action by the LIC$_{\mathrm{CR}}$ from the start determines the posterior dynamics. 

In Fig.~\ref{fig:figure9}b we plot the maximum average number of photons collected for a cavity mode frequency $\omega_{\mathrm{c}}=$5600 cm$^{-1}$ and for a set of different increasing cavity-matter coupling strengths $\chi$ from 0.02 to 0.16. For the lowest couplings 0.02 and 0.04 the photon yield is approximately symmetric against the change of sign of the Huang-Rhys factor.
In these cases, the regime can be safely called strong-coupling because CR effects are not relevant, and the dynamics is only explained in terms of R and DC couplings, which are the same for positive or negative Huang-Rhys factors. 
From $\chi=0.06$ upward the photon yield displays a clear asymmetry in favor of the positive Huang-Rhys factor. Again, it reflects that we arrive at the ultra-strong coupling regime where the CR terms become very active. This separation between strong and ultrastrong coupling with the value of $\chi$ is in agreement with the typical values $g/\omega_{\mathrm{c}} < 0.1$ (strong) and $>0.1$ (ultra-strong), according to $g$ in Eq.~\eqref{eq:Cavityint}.

This prevalence indicates that the correct path for photon conversion is a first meeting with the R terms, which, aided by DC, optimally launches the WF to the CR terms. The latter eventually are the most important agents for high-photon conversion. 
Thus, certain molecules with a specific arrangement of equilibrium distances between the ground and excited states are more likely to convert photons than others.

\subsection{Non-adiabatic polariton photodynamics in CaH}

\begin{figure} [h!]
    \centering
    \includegraphics[width=0.95\linewidth]{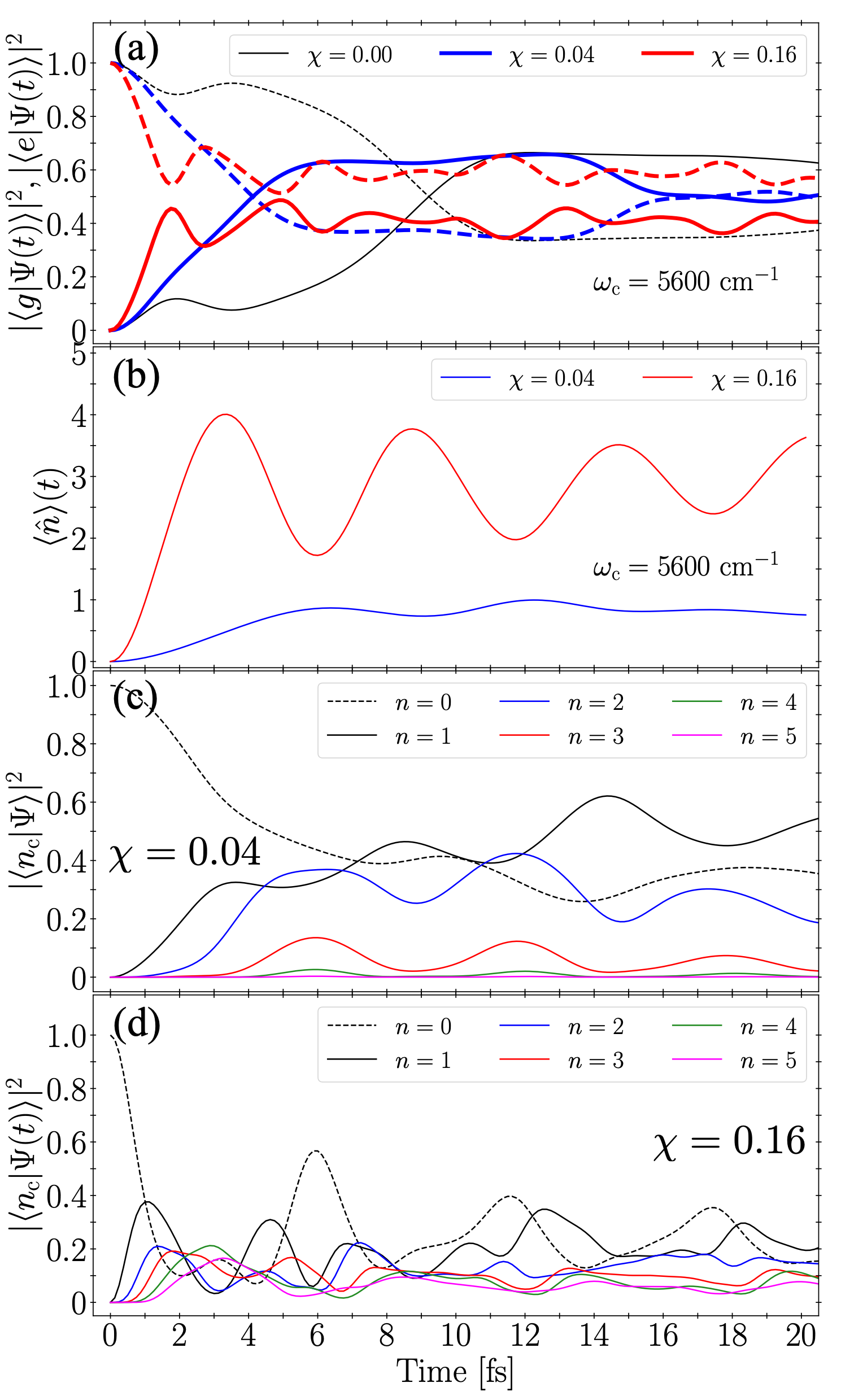}
    \caption{
    (a) Time evolution of the global populations
    $P_g(t) = |\langle g | \Psi(t) \rangle |^2$ (solid lines) and $P_e(t) = |\langle e | \Psi(t) \rangle |^2$ (dashed lines), irrespective of the number of photons, for three cavity coupling strenghts $\chi=0$ (only diabatic coupling), $\chi=0.04$ and $\chi=0.16$.
    (b) Time evolution of the average number of photons $\langle\hat{n}\rangle(t)$ for two coupling strengths $\chi=0.04$ and $\chi=0.16$. 
    (c) Excited state projections $| \langle n_{\mathrm{c}}; e | \Psi (t)\rangle|^2$ for the CaH  B$^2\Sigma^+$ state for the molecule-cavity coupling $\chi=0.04$. 
    (d) Same as (b) for a coupling $\chi=0.16$. }
    \label{fig:figure10}
\end{figure}

We conclude this work by computing the polariton photodynamics of the diatomic molecule CaH from first principles. 
This means that we make use of the {\em ab initio} energies and radiative and non-radiative couplings represented in Fig.~\ref{fig:figure4}. 
For the non-adiabatic coupling, we take the adiabatic coupling $A_{ge}({\bf q})$ calculated with the MOLPRO package, and we make a unitary transformation with an angle $\theta(R)$ (see, for example, \cite{Triana2022b}), to obtain diabatic PECs that now show real crossings, the diabatic (electrostatic) coupling $V_{\mathrm{D}}$ and the diagonal ($d_{gg}$ and $d_{ee}$) and the transition ($d_{ge}$) dipole moments expressed in the diabatic basis. 

For a fully realistic {\em ab initio} computation, the diagonal dipole moments should be included too. In this regard, it means that new light-matter couplings (not included in our extended HQR model) arise, namely $[ d_{gg} \ket{g}\bra{g} +  d_{ee} \ket{e}\bra{e} ]
(\hat{a}^\dag + \hat{a})$. These terms imply that a change in the photon number is not accompanied by an exciton transfer, thus it is a counterpoint to the action of the diabatic coupling. 
The photon energy is thus exchanged by roughly degenerate vibrational states that belong to the same electronic state but with different field dressing, $\ket{n_{\mathrm{c}};g/e} \to \ket{n_{\mathrm{c}}\pm 1; g/e}$ and they redistribute the vibrational populations. 
In this respect, they are not subject to be $q-$located at the energy crossings and they effectively act according to their magnitude along the internuclear distance $q$. Interestingly, these diagonal couplings increase the number of dressed states and mechanistic paths in Fig.~\ref{fig:figure7}a. If included in the extended HQR model, individual transitions due to particular couplings are much more difficult to isolate within the same time window.    

Now we discuss the main features of the CaH fast photodynamics, shown in Fig.~\ref{fig:figure10} for the first 20 fs. As a reference, we have also included populations for $\chi=0$ (without field). The DC in CaH is non zero at the equilibrium distance of the ground state (see Fig.~\ref{fig:figure4}) so that the initial $e$-wave packet is suddenly subject to the extended diabatic coupling, until the inversion of population $\ket{e} \xrightarrow{\mathrm{DC}} \ket{g}$ is accomplished at $t=10$ fs. If the light-molecule interaction is activated with a coupling strength $\chi=0.04$ (assumed to be in the borderline between the strong coupling and ultra-strong coupling regime) the $g \leftrightarrow e$ inversion is produced sooner ($t=4$ fs), prior to the DC crossing in Fig.~\ref{fig:figure10}. Naively, this effect can be attributed to the LIC$_{\mathrm{R}}$, which makes the
expected transition $\ket{0;e} \xrightarrow{\mathrm{R}} \ket{1;g}$, thus contributing to the emerging value of $|\braket{1}{\Psi (t)}|^2$ at short times (0-3 fs) in Fig.~\ref{fig:figure10}c. 
We also appreciate a sequential promotion to 2 and 3 photons, mostly in the $g$ state, namely $\ket{1;g}$, $\ket{2;g}$, and $\ket{3;g}$. 
This photon promotion cannot be attributed to the RWA terms in the LIC$_{\mathrm{R}}$, but due to the sudden DC action $\ket{0;e} \xrightarrow{\mathrm{DC}} \ket{0;g}$ and the concomitant diagonal dipole coupling $d_{gg}(q)$ (relevant in this $q-$region as shown in  Fig.~\ref{fig:figure4}), that brings into scene the 
transitions $\ket{0;g} \xrightarrow{d_{gg}} \ket{1;g} \xrightarrow{d_{gg}} \ket{2;g} \xrightarrow{d_{gg}} \ket{3;g}$. Although the average value of the initial energy is 3.3 eV in CaH, the energy width of the initial wave packet may allow contributions of otherwise closed dressed channels $\ket{2;g}$ and $\ket{3;g}$. 
The polaritonic WF that enters the DC crossing already has a complex superposition 
$\ket{\Psi} = \sum_{n_{\mathrm{c}}=0}^3 (\ket{n_{\mathrm{c}};g} + \ket{n_{\mathrm{c}};e})$.
Given that DC coupling produces exciton exchanges of the type $\ket{n_{\mathrm{c}};g} \leftrightarrow \ket{n_{\mathrm{c}};e}$, the superposition barely modifies across the DC crossing, thus the DC effect remains unnoticed in Fig.~\ref{fig:figure10}a.
After DC crossing, the LIC$_{\mathrm{CR}}$ makes small adjustments in the components of the polariton WF with transitions $\ket{0;g} \xrightarrow{\mathrm{CR}} \ket{1;e}$ and $\ket{2;e} \xrightarrow{\mathrm{CR}} \ket{1;g}$, with an enhancement of the component with $n_{\mathrm{c}}=1$ photon. However, the average number of created photons is limited to 1 photon in Fig.~\ref{fig:figure10}b, as expected by the picture of dressed PECs in Fig.~\ref{fig:figure4}. 

This scenario changes dramatically with the higher coupling strength 
$\chi=0.16$ (ultra-strong coupling). 
The global population exchange or inversion $\ket{e} \leftrightarrow \ket{g}$ follows a kind of quasi-harmonic signal in the time series. This periodic behavior is mostly visible in the average number of photons in Fig.~\ref{fig:figure10}b, which is hard to guess out from the individual photon components in Fig.~\ref{fig:figure10}d (note that 
$\langle \hat{n} \rangle (t)$= $\sum_{n_{\mathrm{c}}} n_{\mathrm{c}} |\braket{n_{\mathrm{c}}}{\Psi(t)}|^2$). Each dressed state $\ket{n_{\mathrm{c}};g/e}$ in the sum incorporates a different Rabi frequency.  This oscillatory pattern in the average number of photons represents a constructive interference effect between all dressed state components in the WF $\ket{\Psi(t)}=\sum_{n_{\mathrm{c}}=0}^{5} ( \ket{n_{\mathrm{c}};g} + \ket{n_{\mathrm{c}};e} )$ built up and equalized by the different diagonal and transition dipole couplings as well as by the DC coupling. In contrast to $\chi=0.04$, now the dynamics within the same time window is faster; before 3 fs the dressed states $\ket{n_{\mathrm{c}};g}$ with $n_{\mathrm{c}}=1-5$ are already generated (see Fig.~\ref{fig:figure10}d), whereas for $\chi=0.04$ the formation of $\ket{n_{\mathrm{c}};g}$, with $n_{\mathrm{c}}=1-3$ takes double the time (see Fig.~\ref{fig:figure10}c). In fact, the oscillatory form in $\langle \hat{n} \rangle (t)$ in Fig.~\ref{fig:figure10}b can be approximately deduced from the global populations in  Fig.~\ref{fig:figure10}a. If we set the zero of the $y$-axis at the average $n_{\mathrm{c}}=3$ photons, the nodes of the function $\langle \hat{n} \rangle (t)$ coincide at the times in which the populations $|\braket{g}{\Psi(t)}|^2$ and $|\braket{e}{\Psi(t)}|^2$ show their closest approach to each other. So to say, if we follow diabatically (making real crossings out of avoided crossings in the global populations) the populations, we can draw the same oscillations as in $\langle \hat{n} \rangle (t)$, indicating an alternating prevalence of the global $g$ or $e$ dressed states, regardless the number of photons.

To conclude, a realistic {\em ab initio} dynamic polariton calculation in CaH, shows that molecules in general must display non-adiabatic conversion of vibronic energy into photons extracted from the electromagnetic vacuum, in this case of CaH as much as 1 photon (4 photons) in the strong (ultrastrong) coupling regime.

\section{\label{sec:conclusions} Conclusions}

To closely understand the inner mechanisms in molecular polaritonics of diatomics during the early stages of the onset of cavity-molecule interaction, we propose and implement an extended Hosltein-quantum-Rabi Hamiltonian. 
The molecular model represents a diabatized version of two molecular excited states of the same symmetry whose potential energy curves display avoided crossings. 
The model includes exciton and vibration modes in diatoms, and cavity photons, all on an equal footing. All these modes are eventually coupled through \textit{i)} non-adiabatic couplings in the diabatic picture, and \textit{ii)} transition dipole moments, both interactions with an explicit dependence on the internuclear distance. 
The shift between the equilibrium internuclear distances of the two excited states is attained in the model by choosing a Huang-Rhys factor.  

The polariton photodynamics studied with this extended HQR model has helped us to understand the complexity of the {\em ab initio} results for the fast dynamics in a molecule as CaH.
Realistic calculations from first principles involve polar molecules with low dissociation thresholds, energy curve crossings, permanent and transition dipole moments, and non-adiabatic couplings, which have a complex structure and are mostly delocalized along the internuclear distance. 
This makes photodynamic observables discriminated in the time domain difficult to interpret at first glance. Our HQR model with localized interactions allows us to clearly isolate transitions in a time sequence. 
For instance, we reveal the different roles of the LIC$_{\mathrm{R}}$, where RWA applies, and  LIC$_{\mathrm{CR}}$, where the counter-rotating non-RWA terms dominate when the molecule-cavity strength approaches the ultra-strong coupling regime.

We demonstrate that the passage of the molecular wave packet through the interacting triplet in the ordering LIC$_{\mathrm{R}}$$\to$DC$\to$LIC$_{\mathrm{CR}}$ guarantees the highest conversion of vibronic energy into cavity photons. 
This ordering is present in molecules with positive Huang-Rys factor, as CaH. Other molecules with PECs arranged with negative Huang-Rys factors have unfavored photon creation.  

We study the photodynamics within the first few tens of fs, which allows for a full cycle of the polariton wave packet. 
Nevertheless, it seems to be the most crucial time window that determines the subsequent dynamics of observables. 
The polariton state, after the first cycle, is already a complex superposition of many dressed states in the form $\ket{\Psi(t)}$=$\sum_{v'',v', n_{\mathrm{c}}} ( \ket{n_{\mathrm{c}};g,v''} + \ket{n_{\mathrm{c}};e,v'} )$, created by the LIC and DC interactions during the first cycle. 
In the next cycles, the spread wave packet is mostly ``thermalized'' in its components, with the same dressed states and similar expansion coefficients already present during the first cycle. 
The ulterior effect of couplings is to produce fluctuations on the already achieved average in the equalized wave packet.     

The photon-dressed picture of molecular states predicts that the combination of exciton, photon, and vibration modes may lead to energy degenerations that eventually allow for the molecular wave packet to be temporarily trapped at a particular internuclear distance in states dressed by a high number of photons.This photon-trapping effect is parallel to the molecular bond-hardening caused by intense lasers.    

Polar molecules also display non-vanishing diagonal dipole moments in ground and excited states, in addition to transition dipole moments. We have not included these permanent dipoles in the HQR model, but in the CaH calculations, which certainly bring about new photonic transitions. We have not included the dipole-self energy term $\frac{1}{2}(\chi\hat{{\bf d}} )^2$ consider in the molecule-cavity Pauli-Fierz Hamiltonian model \cite{Rokaj2018} (see also SM in Ref. ~\cite{Triana2019}), because the light-matter interaction in nanocavities is mainly mediated by Coulomb interactions, which is not influenced by the Power-Zienau-Woolley transformation \cite{Feist2020,fregoni2022}. This approach has been successfully implemented in plasmonic environments or nanocavities \cite{Csehi2025} and is discussed in detail in Refs. \cite{Feist2020,Buhmann2013,Fabri2025}.
Also, in our experience, the dipole self-energy tends to cancel out contributions from permanent dipoles \cite{Triana2020sd}, which leads to the results using our extended HQR model.
Finally, we hope that our study may lead to future considerations for the implementation of a non-adiabatic collective Holstein-Dicke model for molecular ensembles.

\begin{acknowledgments}
A.F.L and J.L.S-V acknowledge financial support from Vicerrectoría de Investigación at Universidad de Antioquia, (CODI, Programática Project 2022-5357 and Estrategia de Sostenibilidad), Colombia.
J.F.T. is supported by grant ANID-Fondecyt Iniciaci\'on 11230679. and he also thanks ``N\'ucleo de Investigaci\'on No. 7 UCN-VRIDT 076/2020, N\'ucleo de modelaci\'on y simulaci\'on cient\'ifica (NMSC)'', for computational support in Chile.
\end{acknowledgments}


\section*{Data Availability Statement}
Most numerical results shown in this work can be ob-
tained from the formulas, parameters and public software described in the text. 
The data that support the findings of this study are available from the corresponding author upon reasonable request.


\appendix

\section{Molecular two-state Hamiltonian based on harmonic potentials.}
\label{sec:AppendixA}
We assume two electronic excited molecular states $g$ and $e$ whose potential energy curves are approximated by using 
harmonic potentials with different frequencies $\omega_g$ and $\omega_e$ and equilibrium distances $q_g=0$ and $q_e \ne 0$ and 
a vertical energy separation $\hbar \omega_{ge} = E^{0}_e-E^{0}_g$. We look for the relation between the two vibrational Hamiltonians 
\begin{equation}
\hat{H}_g = E^0_g + \frac{\hat{p}^2}{2M} + \frac{1}{2} M \omega^2_g \hat{q}_g = E^0_g + \hbar \omega_g \left( \hat{b}^\dag_g \hat{b}_g  + \frac{1}{2}  \right),  
\end{equation}
\begin{equation}
\hat{H}_e = E^0_e + \frac{\hat{p}^2}{2M} + \frac{1}{2} M \omega^2_e \hat{q}_e = E^0_e + \hbar \omega_e \left( \hat{b}^\dag_e \hat{b}_e  + \frac{1}{2}  \right). 
\end{equation}
Since the position operator for the state $|e\rangle$ is displaced with respect to the state $|g\rangle$, there is a relation $\hat{q}_e=\hat{q}_g - q_e$, i.e.,
\begin{align}
\sqrt{ \frac{\hbar}{2M\omega_e} }  \left( \hat{b}^\dag_e + \hat{b}_e \right) & =   \sqrt{ \frac{\hbar}{2M\omega_{g}} }  \left( \hat{b}^\dag_g + \hat{b}_g \right) - q_e \nonumber  \\
& = \sqrt{ \frac{\hbar}{2M\omega_{g}}}  \left[  (\hat{b}_g -\lambda)^\dag   + ( \hat{b}_g - \lambda )  \right]
\end{align}
to obtain the relation
\begin{equation}
\hat{b}^\dag_e + \hat{b}_e = \sqrt{ \frac{\omega_e}{\omega_g} }  \left[  (\hat{b}_g -\lambda)^\dag   + ( \hat{b}_g - \lambda )  \right]
\label{eq:ap1}
\end{equation}
where $\lambda=q_e \sqrt{M \omega_g/2\hbar}$ corresponds to an effective shift constant called Huang-Rhys factor. Similarly, since $\hat{p}_g = \hat{p}_e$ we
arrive to the relation
\begin{equation}
\hat{b}^\dag_e - \hat{b}_e = \sqrt{ \frac{\omega_g}{\omega_e} }  \left[  (\hat{b}_g -\lambda)^\dag   - ( \hat{b}_g - \lambda )  \right].
\label{eq:ap2}
\end{equation}
From adding and subtracting Eqs.~ \eqref{eq:ap1} and \eqref{eq:ap2} one obtains the transformation between the algebraic operators for $g$ and $e$
\begin{align}
\hat{b}_e & = \hat{D}_g (\lambda) \left(  \alpha \hat{b}_g  + \beta \hat{b}^\dag_g  \right) \hat{D}^\dag_g (\lambda) \\
\hat{b}^\dag_e & = \hat{D}_g (\lambda) \left(  \alpha \hat{b}^\dag_g  + \beta \hat{b}_g  \right) \hat{D}^\dag_g (\lambda)
\end{align}
where $\alpha= 1/2 (\sqrt{\omega_e/\omega_g}  + \sqrt{\omega_g/\omega_e} )$ and $\beta=1/2 (\sqrt{\omega_e/\omega_g}  - \sqrt{\omega_g/\omega_e} )$
and $\hat{D}_g (\lambda)= \exp \left[  \lambda ( \hat{b}^\dag_g  - \hat{b}_g  ) \right]$ is the displacement operator for the Huang-Rhys factor $\lambda$. 
These transformations correspond to a squeezing operation if we make the asignment $\alpha= \cosh r$ and $\beta = \sinh r$, where $r= \log ( \sqrt{\omega_e/\omega_g })$ is the squeezing factor. Thus by introducing the squeezing operator $\hat{S}_g (r) = \exp \left[ r (\hat{b}^2_g - \hat{b}^{\dag 2}_g ) /2  \right]$
we arrive to the transformation between the $g$ and $e$ Hamiltonians with
\begin{equation}
\hat{b}^\dag_e \hat{b}_e  = \hat{D}_g (\lambda) \hat{S}_g (r) \hat{b}^\dag_g \hat{b}_g  \hat{S}^\dag_g (r) \hat{D}^\dag_g (\lambda). 
\end{equation} 
Therefore, the eigenstates of the number operator $\hat{b}^\dag_e \hat{b}_e | \tilde{n} \rangle =  n | \tilde{n} \rangle $ for oscillator $e$ are obtained by displacing and compressing those of 
$\hat{b}^\dag_g \hat{b}_g | n \rangle = n | n \rangle$, i.e., $ | \tilde{n} \rangle = \hat{D}_g (\lambda) \hat{S}_g (r) | n\rangle $. We remark that both basis $\left\{ | n \rangle \right\}$ and 
$\left\{ | \tilde{n} \rangle  \right\} $ are complete and both can solve the Schr\"odinger equation for Hamiltonians $\hat{H}_g$ and $\hat{H}_e$. 

\section{Another form for the Holstein-Quantum-Rabi Hamiltonian.}
\label{sec:AppendixB}
It is worth noting that our extended Holstein-Quantum-Rabi Hamiltonian for two diabatic- and dipole-coupled states $g$ and $e$, that can be rewritten in the form
\begin{align}
\hat{H} & =\omega_{\rm c} \hat{a}^\dag \hat{a} + \omega_g \left(    \hat{b}^\dag_g  \hat{b}_g +  \frac{1}{2} \right) \hat{\sigma}^-  \hat{\sigma^+} \nonumber \\ 
             & + \left[   \omega_{ge} + \omega_e \left(  \hat{b}^\dag_e  \hat{b}_e + \frac{1}{2} \right)  \right] \hat{\sigma}^+  \hat{\sigma}^- \nonumber \\
             & + V( \hat{q} ) \left(  \hat{\sigma}^+ +  \hat{\sigma}^-    \right) + g(\hat{q)}      \left(  \hat{\sigma}^+ +  \hat{\sigma}^-    \right) \left(  \hat{a}^\dag +  \hat{a}  \right) 
 \end{align}
 This Hamiltonian can be solved variationally in terms of the complete uncoupled basis sets $|n_{\rm c}; g, n \rangle$ and shifted-squeezed $|n_{\rm c}; e, \bar{n} \rangle$. We can transform the basis and the Hamiltonian with a dilation-displacement unitary operator $\hat{U} = \hat{1} + \hat{\sigma}^+  \hat{\sigma}^- \left[    \hat{S}^\dag(r) \hat{D}^\dag (\lambda)    \right] $. Using this transformation we find $\hat{U} |n_{\mathrm{c}}; g,n \rangle =  |n_{\rm c}; g, n \rangle$ and $\hat{U} |n_{\rm c}; e, \bar{n} \rangle =  |n_{\rm c}; e, n \rangle$, that implies that the vibrational set $|n\rangle$ is now the same for $g$ and $e$ states. It is straightforward to show that radiation operator remains invariant under the transformation but matter operators are transformed as follows
\begin{enumerate}
\item exciton operators            
\begin{align}
                \hat{U} \hat{\sigma}^{+} \hat{U}^{\dagger} &= \hat{S}^{\dagger}(r) \hat{D}^{\dagger}(\lambda) \hat{\sigma}^{+}, \nonumber \\
                \hat{U} \hat{\sigma}^{-} \hat{U}^{\dagger} &=  \hat{D}(\lambda) \hat{S}(r) \hat{\sigma}^{-}, \nonumber \\
                \hat{U} \hat{\sigma}^{+} \hat{\sigma}^{-} \hat{U}^{\dagger} &= \hat{\sigma}^{+} \hat{\sigma}^{-} \nonumber,\\
                \hat{U} \hat{\sigma}^{-} \hat{\sigma}^{+} \hat{U}^{\dagger} &= \hat{\sigma}^{-} \hat{\sigma}^{+}.
\end{align}
\item phonon operators in state $g$
\begin{align}
& \hat{U} \hat{b}_g\, \hat{U}^{\dag} = \hat{b}_g  \nonumber \\ 
& \qquad  + \hat{\sigma}^{+} \hat{\sigma}^{-} \left[ \hat{S}^{\dag}(r) \hat{D}^{\dag}(\lambda) \hat{b}_g \hat{D}(\lambda) \hat{S}(r) - \hat{b}_g \right ], \nonumber \\
&\hat{U} \hat{b}_g^{\dag} \hat{U}^{\dag} =  \hat{b_g}^{\dag} \nonumber \\ 
&\qquad + \hat{\sigma}^{+}\hat{\sigma}^{-} \left[ \hat{S}^{\dag}(r) \hat{D}^{\dag}(\lambda) \,\hat{b}_g^{\dag} \hat{D}(\lambda) \hat{S}(r) - \hat{b}_g^{\dag}  \right], \nonumber \\
&\hat{U} \hat{b}_g^{\dag} \hat{b}_g  \hat{U}^{\dag}  = \hat{b}_g^{\dag}\hat{b}_g \nonumber \\ 
& \qquad + \hat{\sigma}^{+} \hat{\sigma}^{-} \left[ \hat{S}^{\dag}(r) \hat{D}^{\dag}(\lambda) \hat{b}_g^{\dag} \hat{b}_g  \hat{D}(\lambda) \hat{S}(r) - \hat{b}_g^{\dag}\hat{b}_g \right].
\end{align}
\item phonon operators in state $e$, for the anhilitation operator $\hat{b}_e$ = $\hat{D}(\lambda)\hat{S}(r)\,\hat{b}\,\hat{S}^{\dagger}(r)\hat{D}^{\dagger}(\lambda)$:
\begin{align}
\hat{U} \hat{b}_e \hat{U}^{\dag} &= \hat{b}_e + \hat{\sigma}^{+} \hat{\sigma}^{-} \left[ \hat{b}_g - \hat{b}_e \right], \nonumber \\
\hat{U} \hat{b}^{\dag}_e \hat{U}^{\dag} &=\hat{b}^{\dagger}_e + \hat{\sigma}^{+} \hat{\sigma}^{-} \left[  \hat{b}_g^{\dag} - \hat{b}^{\dag}_e \right], \nonumber \\
\hat{U} \hat{b}^{\dag}_e \hat{b}_e \hat{U}^{\dag} &= \hat{b}^{\dag}_e \hat{b}_e + \hat{\sigma}^{+} \hat{\sigma}^{-} \left[ \hat{b}_g^{\dag} \hat{b}_g - \hat{b}^{\dag}_e\hat{b}_e \right].
 \end{align}
\end{enumerate}     
Using the previous transformations, the transformed Hamiltonian now reads
    \begin{align}
        \hat{U} & \hat{H} \hat{U}^{\dag}  = \omega_{\rm c} \hat{a}^{\dag}\hat{a} + \omega_{ge}\hat{\sigma}^{+} \hat{\sigma}^{-}  \nonumber \\
        &  + \left( \omega_g \hat{\sigma}^{-} \hat{\sigma}^{+} + \omega_e \hat{\sigma}^{+} \hat{\sigma}^{-} \right)  \left (\hat{b}_g^{\dag} \hat{b}_g + \frac{1}{2} \right) \nonumber \\
        &  + V(\hat{q}) \left[  \hat{S}^{\dag}(r) \hat{D}^{\dag}(\lambda) \hat{\sigma}^{+}  + \hat{D}(\lambda)\hat{S}(r) \hat{\sigma}^{-} \right]  \nonumber \\
        &  + g(\hat{q}) \left[ \hat{S}^{\dag}(r) \hat{D}^{\dag}(\lambda)\hat{\sigma}^{+} + \hat{D}(\lambda)\hat{S}(r)\hat{\sigma}^{-} \right] \left( \hat{a}^{\dag} + \hat{a} \right),
    \end{align}   
which involves the calculation of matrix elements involving the basis states $\ket{n_\mathrm{c};g/e,n}$.

\bibliographystyle{apsrev4-1}
\bibliography{bibliography}

\end{document}